\documentclass[reprint,amsmath,amssymbGaps,onecolumn,notitlepage,nofootinbib]{revtex4-1}

\usepackage{amsmath} 
\usepackage{amssymb}
\usepackage{amsfonts}
\usepackage{amstext}
\usepackage{graphicx}
\usepackage{bm}
\usepackage{color} 
\usepackage{transparent}
\usepackage{mathtools}
\usepackage{slashed}
\usepackage{hyperref}
\usepackage{multirow}
\usepackage{textgreek}

\newcommand{\cbr}[1]{\left(#1\right)}
\newcommand{\sbr}[1]{\left[#1\right]}

\newcommand{\boxp}{\Box \phi}
\newcommand{\dpp}{\left(\nabla \phi \right)^2}

\newcommand{\GB}{\mathcal{G}}

\begin{document}

\title{Exploring the Small Mass Limit of Stationary Black Holes in Theories with Gauss-Bonnet Terms}
\author{Pedro G. S. Fernandes$^{1,2}$}
\email{pedro.fernandes@nottingham.ac.uk}
\author{David J. Mulryne$^{1}$}
\author{Jorge F. M. Delgado$^{3,4}$}
\affiliation{$^{1}$School of Physics and Astronomy, Queen Mary University of London, Mile End Road, London, E1 4NS, UK}
\affiliation{$^{2}$School of Physics and Astronomy, University of Nottingham, University Park, Nottingham, NG7 2RD, United Kingdom}
\affiliation{$^{3}$Departamento de Matemática da Universidade de Aveiro and Centre for Research and Development in Mathematics and Applications (CIDMA)\\Campus de Santiago, 3810-183 Aveiro, Portugal\\}
\affiliation{$^{4}$Centro de Astrofísica e Gravitação - CENTRA,\\
Departamento de Física, Instituto Superior Técnico - IST, Universidade de Lisboa - UL,\\
Av. Rovisco Pais 1, 1049-001 Lisboa, Portugal}


\begin{abstract}

\par In this work we examine the small mass limit of black holes, with and without spin, in theories where a scalar field is non-minimally coupled to a Gauss-Bonnet term. First, we provide an analytical example for a theory where a static closed-form solution with a small mass limit is known, and later use analytical and numerical techniques to explore this limit in standard scalar-Gauss-Bonnet theories with dilatonic, linear and quadratic-exponential couplings. In most cases studied here, we find an inner singularity that overlaps with the event horizon of the static black hole as the small mass limit is reached. Moreover, since solutions in this limit possess a non-vanishing Hawking temperature, a naked singularity is expected to be reached through evaporation, raising questions concerning the consistency of these theories altogether. On the other hand, we provide for the first time in this context an example of a coupling where the small mass limit is never reached, thus preferred from the point of view of cosmic censorship. Finally, we consider black holes with spin and numerically investigate how this changes the picture, using these to place the tightest upper bounds to date on the coupling constant for the dilatonic and linear theories, with  $\sqrt{\overline{\alpha}} < 1$ km.

\end{abstract}

\maketitle

\section{Introduction} \label{intro}

Prominent examples of alternative theories of gravity to General Relativity (GR) are \textit{Einstein-scalar-Gauss-Bonnet} (EsGB) theories
, where a new fundamental scalar is non-minimally coupled to the Gauss-Bonnet (GB) term
\begin{equation}
    \mathcal{G} = R^2 - 4 R_{\mu \nu}R^{\mu \nu} + R_{\mu \nu \alpha \beta} R^{\mu \nu \alpha \beta}.
\end{equation}
Such models belong to the Horndeski class of theories \cite{Horndeski:1974wa, Kobayashi:2019hrl}, and in the simplest case their action takes the form
\begin{equation}
    S=\frac{1}{16\pi} \int d^4x \sqrt{-g} \cbr{R-\dpp +\frac{\alpha}{4} \,\xi\cbr{\phi} \GB},
    \label{eq:action}
\end{equation}
where $\phi$ is a real scalar field, $\xi\cbr{\phi}$ is the (non-minimal) coupling function, and $\alpha$ the GB coupling constant with dimensions of length squared.

EsGB theories are of wide interest and have been the subject of many works in recent years (see e.g.  \cite{Sotiriou:2013qea,Sotiriou:2014pfa,Delgado:2020rev,Doneva:2017bvd,Silva:2017uqg,Antoniou:2017acq,Macedo:2019sem,Saravani:2019xwx,Cunha:2019dwb,Dima:2020yac,Doneva:2020nbb,Herdeiro:2020wei,Berti:2020kgk,Herdeiro:2021vjo,Doneva:2021tvn,Kanti:1995vq,Kleihaus:2011tg,Cunha:2016wzk, Andreou:2019ikc,Franchini:2022ukz,Ripley:2019hxt,Ripley:2019irj,Oikonomou:2021kql,Oikonomou:2020sij,Odintsov:2020sqy,Maeda:2009uy}). From a fundamental physics perspective they arise, for example, as the low-energy limit of some string theories \cite{Nepomechie:1985us,Gross:1986mw,Candelas:1985en,Callan:1986jb} where the scalar, \textit{the dilaton}, couples exponentially to the GB term, $\xi \cbr{\phi} \sim e^{\gamma \phi}$ \cite{Kanti:1995vq,Kleihaus:2011tg,Cunha:2016wzk,Herdeiro:2018wub,Maeda:2009uy}. From a more phenomenological point of view, they are one 
of a variety of theories \cite{Sotiriou:2013qea,Sotiriou:2014pfa,Delgado:2020rev,Doneva:2017bvd,Silva:2017uqg,Antoniou:2017acq,Macedo:2019sem,Saravani:2019xwx,Cunha:2019dwb,Dima:2020yac,Doneva:2020nbb,Herdeiro:2020wei,Berti:2020kgk,Herdeiro:2021vjo,Kanti:1995vq,Kleihaus:2011tg,Cunha:2016wzk,Herdeiro:2018wub,Fernandes:2019rez,Fernandes:2019kmh,Fernandes:2020gay} that 
evade the no hair conjecture \cite{Ruffini1971,Bekenstein:1996pn, PhysRevD.5.1239,Sotiriou:2015pka,Hui:2012qt,Bekenstein:1995un,Hawking:1972qk,Sotiriou:2011dz} (see \cite{Herdeiro:2015waa} for a review), raising the exciting possibility they can be constrained by black hole (BH) physics in the strong-curvature regime.
Moreover, in some EsGB theories 
a dynamical mechanism called \textit{spontaneous scalarization} \cite{Doneva:2017bvd,Silva:2017uqg,Antoniou:2017acq,Dima:2020yac,Herdeiro:2020wei,Berti:2020kgk,Herdeiro:2018wub,Silva:2018qhn,Blazquez-Salcedo:2018jnn,Fernandes:2019rez,Fernandes:2019kmh,Fernandes:2020gay} can occur, such that deviations from GR occur \textit{only} in the strong-curvature regime. Constraints can be theoretical, such as self-consistency, or observational in nature. Indeed, only recently has GR begun to be tested observationally in the strong-field regime \cite{Berti:2015itd}, with the dawn of gravitational wave (GW) astronomy \cite{LIGOScientific:2016aoc,LIGOScientific:2017vwq,LIGOScientific:2020ibl}.

\par In a striking difference to GR, some EsGB black holes (e.g. with linear and dilatonic couplings) are known to possess a minimum mass solution whose Hawking temperature is finite and non-vanishing, naturally raising the question of what is the fate of black holes in EsGB theories (see e.g. \cite{Kanti:1995vq,Sotiriou:2014pfa,Torii:1996yi,Alexeev:1996vs,Pani:2009wy,Guo:2008hf,Alexeyev:2009kx,Maeda:2009uy}). This conundrum is often overlooked, but was recently explored in Refs. \cite{Corelli:2022pio,Corelli:2022phw}, where non-linear numerical simulations of evaporating EsGB dilatonic black holes were performed, supporting the idea that the end-point of Hawking evaporation is likely a naked singularity, violating weak cosmic censorship.  This is a rather concerning scenario, raising questions about the consistency of EsGB models.

\par In this work our main purpose is to explore further the small mass limit of black holes for several couplings in EsGB theories, including those allowing for spontaneous scalarization, where a detailed analysis of the small mass limit is so far lacking. Our aim is to investigate self-consistency and observational constraints imposed on Gauss-Bonnet theories and their coupling dependence. After providing a novel example of a closed-form solution with a small mass limit, we will complement previous studies \cite{Kanti:1995vq,Sotiriou:2014pfa,Torii:1996yi,Alexeev:1996vs,Pani:2009wy,Corelli:2022pio,Corelli:2022phw,Alexeyev:2009kx,Maeda:2009uy} on EsGB black hole solutions with a thorough analytical and numerical exploration of the domain of existence of solutions and their inner structure, linking the existence of an inner singularity to the repulsive effects originating from the Gauss-Bonnet term, and to the structure of the field equations. Using analytical arguments, this singularity will be shown to overlap with the event horizon in the small mass limit. We will also provide for the first time, in this context, an example of a coupling function (quadratic-exponential coupling in Eq. \eqref{eq:coupling-quad} below, with $\beta$ above a certain value) where the small mass limit is never reached, showing that a minimum size solution is \textit{not} a generic feature of EsGB theories\footnote{Ref. \cite{Doneva:2017bvd} explores a similar coupling. However, in Ref. \cite{Doneva:2017bvd} no comment is made on the implications for the small mass limit of black holes within the theory. Rather, the authors focus on the domain of existence of scalarized black hole solutions, which they follow all the way to vanishing masses.}. Finally, we construct stationary black hole solutions by numerically solving the field equations in axi-symmetry, with the aim of exploring the small mass limit once spin is considered, and finish by imposing the tightest constraints on the coupling constant, to date, on the dilatonic and linear theories.

\par The rest of the paper is organized as follows. First, in Section \ref{sec:4DEGB}, we introduce a theory which, unlike Eq.~\eqref{eq:action}, admits a known analytical example of a static black hole with a small mass limit. This allows us to explore key features with the advantage of an exact solution. In Section \ref{sec:shape_coupling}, we discuss the form of coupling functions in standard Gauss-Bonnet theories and their corresponding different phenomenologies. Then in Section \ref{sec:static}, we explore the small mass limit of static black hole solutions for these theories, and later impose upper bounds on the coupling constant $\alpha$ in Section \ref{sec:constraints}. Finally, in Section \ref{sec:spin}, we consider how spin changes the picture. We conclude in Section \ref{conclusions}.

\section{Exploring the small mass limit: An analytical example}
\label{sec:4DEGB}
\par No analytic closed-form black hole solution is known to EsGB models described by the action \eqref{eq:action}. In the existing literature, therefore, the study of critical solutions in such theories  
has been performed by resorting to numerical techniques. In this section, we will explore an illustrative example of a related theory with known closed-form black hole solutions.

\par
\begin{figure}[h!]
\centering
\includegraphics[width=0.5\textwidth]{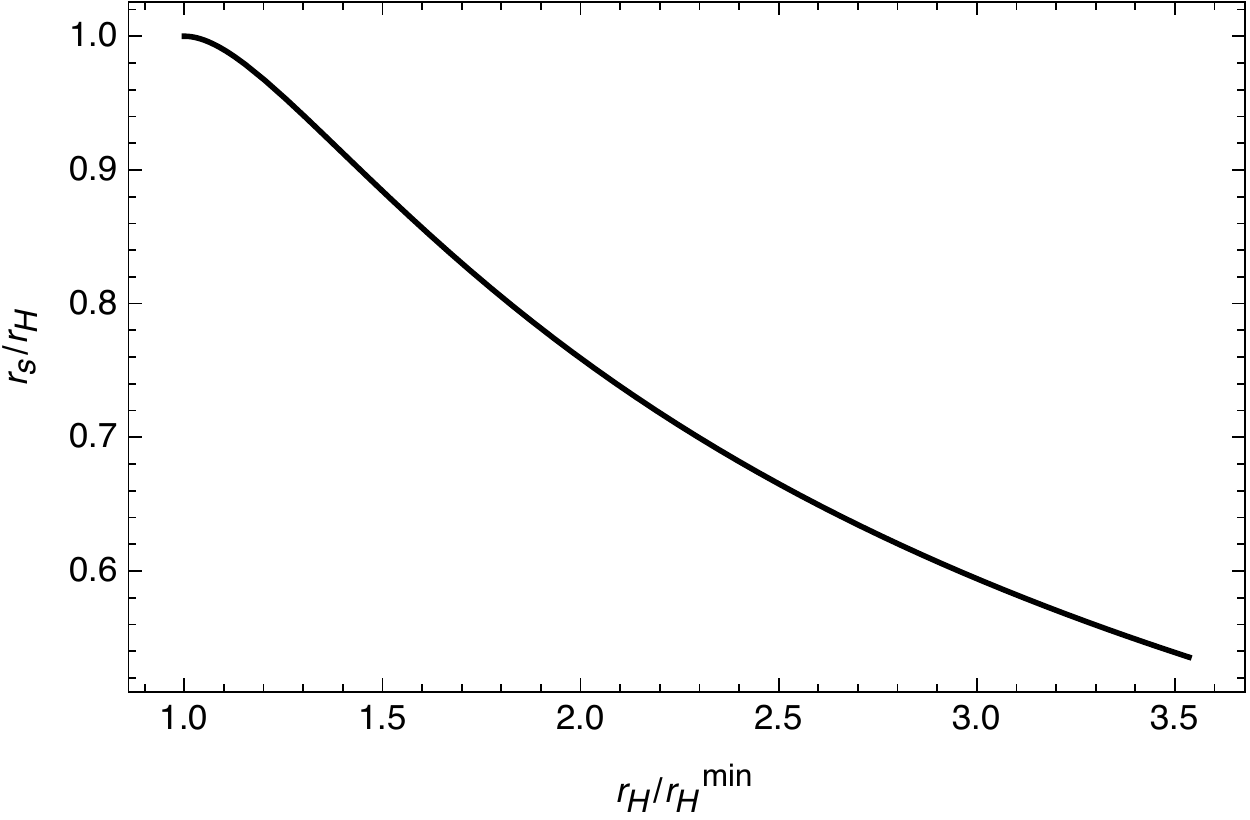}
\caption{Location of the finite radius singularity $r_s/r_H$ as a function of $r_H/r_H^{min}$ for the black hole solution of Eq. \eqref{eq:4degb_sol}. We observe that $r_s$ and $r_H$ overlap as $r_H \to r_H^{min}$.}
\label{fig:rsrh4degb}
\end{figure}

 The theory is known as \textit{gravity with a generalized conformal scalar field}, and was first derived in Ref. \cite{Fernandes:2021dsb} by imposing conformal invariance on the equation of motion of the scalar field. Its action is given by
\begin{equation}
        S = \int \frac{d^{4} x \sqrt{-g}}{16\pi} \bigg[R - \left(\nabla \Phi\right)^2 - \frac{R}{6}\Phi^2  + \alpha \bigg(\log(\Phi) \mathcal{G} - \frac{4G^{\mu \nu}\nabla_{\mu} \Phi \nabla_{\nu} \Phi}{\Phi^2} - \frac{4\square \Phi(\nabla \Phi)^{2}}{\Phi^3} + \frac{2(\nabla \Phi)^{4}}{\Phi^4}\bigg)+ \frac{\lambda}{4} \Phi^4\bigg],
    \label{eq:confcoupled}
\end{equation}
which contains the terms present in the EsGB model of Eq. \eqref{eq:action} with a logarithmic coupling, as well as other non-trivial interactions including a conformal coupling to gravity and a conformally invariant quartic self-coupling. The above theory is intimately connected with the scalar-tensor formulations of so called 4D-Einstein-Gauss-Bonnet gravity \cite{Glavan:2019inb,Lu:2020iav,Kobayashi:2020wqy,Fernandes:2020nbq,Hennigar:2020lsl} (see Ref. \cite{Fernandes:2022zrq} for a review), and belongs to the Horndeski class with functions specified by \cite{Horndeski:1974wa, Kobayashi:2019hrl}
\begin{equation}
    \begin{aligned}
        &G_2 = 2X + \frac{\lambda}{4}\Phi^4 + \frac{8 \alpha X^2}{\Phi^4} \left(6\log X - 23\right), \qquad G_3 = -\frac{24 \alpha X}{\Phi^3} \left(\log X - 3\right),\\&
        G_4 = 1-\frac{\Phi^2}{6}+\frac{4\alpha X}{\Phi^2} \left(\log X - 3\right), \qquad G_5 = -\frac{4\alpha}{\Phi} \log X, \qquad X \equiv -\frac{1}{2}\left(\nabla \Phi\right)^2.
    \end{aligned}
\end{equation}
Note that for the purpose of presentation, we have flipped the sign of $\alpha$ relative to the presentation of Ref. \cite{Fernandes:2021dsb}, and that of typical 4D-Einstein-Gauss-Bonnet studies. Among the many interesting features of Eq. \eqref{eq:confcoupled}, one that stands out is that a special combination of the field equations decouples from the scalar field, imposing a proportionality condition between the Ricci and GB scalars
\begin{equation}
    R=\frac{\alpha}{2} \mathcal{G}\,.
\end{equation}
This allows for an easy search of closed form solutions. One known closed-form black hole solution to the above theory (with $\lambda^{-1} = 6\alpha$) is given by \cite{Fernandes:2021dsb}\footnote{See also Ref. \cite{Babichev:2022awg} for a discussion of other solutions of the theory.}
\begin{equation}
ds^2 = -f(r) e^{-2\delta(r)}dt^2 + \frac{dr^2}{f(r)} + r^2 \cbr{d\theta^2 + \sin^2 \theta d \varphi^2},
\label{eq:lineelement}
\end{equation}
with
\begin{equation}
    \begin{aligned}
        f(r) = 1-\frac{r^2}{2\alpha} \left(1-\sqrt{1-\frac{8 M \alpha}{r^3}}\right), \qquad \delta(r) = 0, \qquad
        \Phi(r) = \frac{2\sqrt{3\alpha}}{r} \mathrm{sech} \left(\int^r \frac{dr}{r\sqrt{f}} \right),
    \end{aligned}
    \label{eq:4degb_sol}
\end{equation}
where $M$ is the ADM mass of the black hole, and we take $\alpha$ to be positive. The scalar field can be seen to be regular on and outside the event horizon located at
\begin{equation}
    r_H = M + \sqrt{M^2+\alpha}.
    \label{eq:4degb_horizon}
\end{equation}
Analysing the Ricci scalar of the solution \eqref{eq:4degb_sol} we observe that
\begin{equation}
    R \propto r^{-3/2} \cbr{r^3 - 8 M \alpha}^{-3/2},
\end{equation}
revealing the existence of two physical singularities, one located at $r=0$, and a finite radius singularity located at the point where the quantity inside the square-root in Eq. \eqref{eq:4degb_sol} for the function $f(r)$ vanishes
\begin{equation}
    r = r_s = 2 \cbr{M \alpha}^{1/3} > 0.
    \label{eq:4degb_sing}
\end{equation}
To ensure physical behaviour of the solution we require that i) the singularity located at $r=r_s$ is hidden behind the event horizon ($r_s < r_H$); ii) the metric functions and the scalar field in Eq. \eqref{eq:4degb_sol} are real.
Under these requirements, it can be shown using Eqs. \eqref{eq:4degb_sol}, \eqref{eq:4degb_horizon}, and \eqref{eq:4degb_sing} that the following condition must hold
\begin{equation}
    \frac{M}{\sqrt{\alpha}} > \frac{1}{2\sqrt{2}} \approx 0.353553\,,
\end{equation}
or, in terms of $r_H$,
\begin{equation}
    \frac{r_H}{\sqrt{\alpha}} > \sqrt{2} \approx 1.41421\,.
\end{equation}
In other words there is a minimum mass $M^{min} = \frac{\sqrt{\alpha}}{2\sqrt{2}}$ (or equivalently, a minimum horizon radius $r_H^{min} = \sqrt{2\alpha}$), below which solutions can no longer be described by black holes. For an object with $r_H = r_H^{min}$, $r_s$ and $r_H$ overlap, as can be observed in Fig. \ref{fig:rsrh4degb}.
A possible physical interpretation for the minimum mass solution is related to the repulsive effect of the Gauss-Bonnet term on the solutions. Examining the components of the effective stress-energy tensor we get
\begin{equation}
    \rho \equiv T^{t}_{\phantom{t}t, eff} = -p_r, \qquad p_r \equiv T^{r}_{\phantom{r}r,eff} = \frac{3 \left(1-\sqrt{1-8 \alpha  M/r^3}\right)^2}{4 \alpha  \sqrt{1-8 \alpha  M/r^3}},
\end{equation}
where $\rho$ and $p_r$ are interpreted as the effective energy density and radial pressure, respectively.\footnote{Note that the dominant energy condition is saturated and e.g. the weak energy condition is violated.} The effective radial pressure is positive everywhere (repulsive), and diverges at $r=r_s$. Indeed, for the minimum mass solution, the repulsive effects of the Gauss-Bonnet term dominate over the standard attractive ones, impeding the existence of a regular horizon.

\par An interesting remark can be made regarding the Hawking temperature of the black hole solution given by
\begin{equation}
    T_H = \frac{1}{4\pi} f'(r_H) = \frac{r_H^2+\alpha}{4\pi r_H \left(r_H^2-2\alpha \right) }\,.
\end{equation}
This is that not only is the temperature non-zero as $r_H \to r_H^{min}$, but in fact diverges! Therefore, as the small mass limit is approached, evaporation will not halt and the black hole will continue to lose its mass at a rate \cite{doi:10.1142/1301}
\begin{equation}
    \frac{dM}{dt} = -\frac{1}{2\pi} \sum_{\ell, m} \int d\omega \frac{\omega G_{\ell m}(\omega)}{e^{\omega/T_H}\pm 1},
    \label{eq:greybody}
\end{equation}
where $G_{\ell m}(\omega)$ are the graybody factors for modes with frequency $\omega$, angular dependence $(\ell, m)$, and the plus/minus sign is related to the emission of fermions/bosons. This intriguing feature casts doubts on the endpoint of Hawking evaporation, with the above calculations suggesting that a naked singularity is a strong endpoint candidate.

\par To summarize, there are a few lessons to be learned from the example. First, it appears that theories with higher-curvature terms are susceptible to the existence of a finite radius singularity located at $r_s>0$. The existence of this singularity is, mathematically, intimately tied to terms containing square-roots in the solutions to the field equations, and the requirement that solutions are real. From a more physical point of view, the singularity is related to repulsive effects originating from the Gauss-Bonnet term. Secondly, a minimum mass solution might exist in these kinds of theories, where the location of the finite radius singularity and of the event horizon overlap. In the above example, this \textit{critical} solution possesses a non-vanishing Hawking temperature, presumably leading to the formation of a naked singularity.

\section{Einstein-scalar-Gauss-Bonnet gravity: Field Equations and the shape of $\xi \cbr{\phi}$}
\label{sec:shape_coupling}
\par We now consider the more standard framework of the action of Eq. \eqref{eq:action}. Varying with respect to the metric tensor we obtain the Einstein equations
\begin{equation}
\mathcal{E}_{\mu \nu} \equiv G_{\mu \nu} - T_{\mu \nu} = 0,
\end{equation}
where
\begin{equation*}
    T_{\mu \nu} = \nabla_\mu \phi \nabla_\nu \phi - \frac{1}{2} g_{\mu \nu} \dpp + \alpha\, ^{\star}R^{\star}_{\mu \alpha \nu \beta} \nabla^\alpha \nabla^\beta \xi\cbr{\phi},
\end{equation*}
and
\begin{equation*}
        ^{\star}R^{\star}_{\alpha \beta \mu \nu} \equiv \frac{1}{4} \epsilon_{\alpha \beta \gamma \delta} R^{\rho \sigma \gamma \delta} \epsilon_{\rho \sigma \mu \nu} = 2\, g_{\alpha [\mu}G_{\nu] \beta} + 2\, g_{\beta [\nu} R_{\mu] \alpha} -R_{\alpha \beta \mu \nu},
\end{equation*}
is the double-dual Riemann tensor (the square brackets denote anti-symmetrization). The scalar field equation is
\begin{equation}
\mathcal{E}_{\phi} \equiv \boxp + \frac{\alpha}{8} \dot \xi(\phi) \GB = 0,
\label{eq:sfequation}
\end{equation}
where the dot denotes differentiation with respect to the scalar field $\phi$.

Starting with the scalar field equation \eqref{eq:sfequation}, we review how different shapes of the coupling function $\xi\cbr{\phi}$ allow different phenomenologies. 
First we note that classical vacuum GR solutions require that $\phi=0$, and that these solutions only exists for couplings that obey the condition $\dot \xi(0) = 0$. On the other hand solutions of models whose couplings obey $\dot \xi(0) \neq 0$ necessarily differ from those of GR and possess a non-trivial scalar field.
Common examples of couplings obeying $\dot \xi(0) \neq 0$ are
\begin{equation}
\xi(\phi) = e^{\gamma \phi},
\label{eq:coupling-exp}
\end{equation}
and
\begin{equation}
\xi(\phi) = \phi.
\label{eq:coupling-linear}
\end{equation}
The first, hereby dubbed the \textit{dilatonic} (or exponential) coupling, is motivated from string theory, as it is the coupling that appears in the 4D low-energy limit of heterotic string theory \cite{Nepomechie:1985us,Gross:1986mw,Candelas:1985en,Callan:1986jb,Kanti:1995vq,Kleihaus:2011tg,Cunha:2016wzk,Herdeiro:2018wub}. The second -- the linear coupling -- can be considered as a linearization of the first around $\phi=0$, and additionally possesses a shift-symmetry in the scalar field 
\cite{Sotiriou:2013qea,Sotiriou:2014pfa,Delgado:2020rev}. For the dilatonic coupling, we focus on the $\gamma=1$ case. Note, however, that as discussed in \cite{Corelli:2022phw,Corelli:2022pio,Guo:2008hf} the properties of black hole solutions might differ for other values of $\gamma$. In fact, for sufficiently negative values ($\gamma \lesssim -1$), the small mass limit and the small size (radius) limit might not coincide, with small mass limit solutions being regular on and outside the horizon. For all the solutions discussed on this work, however, the minimum size and mass limits coincide, and therefore we will use these terms interchangeably. We chose to focus on the $\gamma=1$ case because values $\gamma \lesssim -1$ were explored in Refs. \cite{Corelli:2022phw,Corelli:2022pio,Guo:2008hf}, and in our numerical explorations, all other values of $\gamma$ to lead to similar behaviour as the one we observe for $\gamma=1$. Therefore, we choose $\gamma=1$ as a fiducial value.

As discussed, if the coupling function satisfies $\dot \xi(0)=0$, then $\phi=0$ solves the field equations and the GR solutions are solutions to the model. If however, $\ddot \xi(0) > 0$ then the GR solutions are subject to tachyonic instabilities in the large curvature regime. This can be seen by linearizing the scalar field equation around GR solutions, for example about the Schwarzschild solution. For a perturbation $\delta \phi$ (for which $\dot \xi(0+\delta \phi) \approx \dot \xi(0) + \ddot \xi(0) \delta \phi$) one finds that 
\begin{equation}
\left(\square_{GR} + \frac{\alpha}{8} \ddot \xi(0) \mathcal{G}_{GR}\right) \delta \phi \equiv \left(\square_{GR} - \mu_{eff}^2 \right) \delta \phi = 0,
\label{eq:perturbation}
\end{equation}
where $\mathcal{G}_{GR} = 48M^2/r^6 =12r_H^2/r^6>0$ (with $M$ being the ADM mass of the black hole and $r_H$ the event horizon radius in Schwarzschild-like coordinates). Thus for $\alpha >0$, if $\ddot \xi(0) >0$ the Schwarzschild black hole may develop an instability (as the effective mass gets negative, $\mu_{eff}^2 <0$) \footnote{See also Ref. \cite{Doneva:2021tvn} for the special case where $\ddot \xi(0) = 0$. In this situation, no tachyonic instability exists but GR black holes may become unstable against non-linear scalar perturbations, leading to the formation of scalarized black holes.}. In this tachyonic regime, it has been shown that another class of solutions with a non-trivial scalar-field profile coexists with the GR solutions and are dynamically preferred, triggering spontaneous scalarization. In order to explore spontaneous scalarization we assume couplings of these type to obey the conditions
\begin{equation}
\xi(0) = 0, \qquad \dot \xi(0) = 0, \qquad \ddot \xi(0) = 1,
\label{eq:coupling_conditions}
\end{equation}
The first condition can be imposed as the theory is invariant under $\xi(\phi) \to \xi(\phi) + cte$, the second condition arises by requiring the existence of GR solutions and third condition can be imposed without loss of generality while maintaning a tachyonic instability in the large curvature regime. An example of such a coupling, commonly used in the literature \cite{Doneva:2017bvd,Doneva:2021dqn,Staykov:2021dcj,Danchev:2021tew,Blazquez-Salcedo:2020rhf,Blazquez-Salcedo:2018jnn,Herdeiro:2020wei,Cunha:2019dwb}, and that we shall study here is the ``quadratic exponential" coupling 
\begin{equation}
\xi(\phi) = \frac{1}{2\beta} \left(1-e^{-\beta \phi^2} \right),
\label{eq:coupling-quad}
\end{equation}
where $\beta$ is a constant. Note that this choice is by no means unique, and a simple quadratic coupling $\xi(\phi) = \phi^2/2$ (which is a particular case of Eq. \eqref{eq:coupling-quad} in the limit of vanishing $\beta$) would suffice to explore spontaneous scalarization \textit{per se}. However, black hole solutions in models with a simple quadratic coupling are unstable, contrarily to those with a quadratic exponential coupling \cite{Macedo:2019sem}, and phenomenologies might differ. Note that the $\beta=3$ case was studied extensively e.g. in Refs. \cite{Doneva:2017bvd,Blazquez-Salcedo:2018jnn}. We find that for any coupling obeying the conditions \eqref{eq:coupling_conditions} the instability of a Schwarzschild black hole exists for (see Appendix \ref{app:onset} for a detailed discussion)
\begin{equation}
    r_H/\sqrt{\alpha} \lesssim 0.83.
    \label{eq:condition2}
\end{equation}

Finally, we would like to point that another type of scalarization is possible, being induced by spin. For the Kerr metric, while for dimensionless spins $\chi \leq 0.5$ the Gauss-Bonnet term is positive definite, this is no longer true when higher spins are considered \cite{Hod:2020jjy,Hod:2022htt}. Therefore, if along with the other conditions in Eq. \eqref{eq:coupling_conditions} $\ddot \xi(0)$ is negative (instead of positive), fast-spinning Kerr black holes might be subject to a tachyonic instability \cite{Dima:2020yac,Herdeiro:2020wei,Berti:2020kgk}. A coupling compatible with this type of scalarization would be the coupling of Eq. \eqref{eq:coupling-quad} with a reversed overall sign. We will briefly discuss spin-induced scalarized solutions in Sec. \ref{sec:spin}.

\section{Static black hole solutions, and their small mass limit} 
\label{sec:static}

\par In the existing literature, it has been identified that static EsGB black holes exhibit similar behaviour to that described for the analytic case of Section \ref{sec:4DEGB}. 
Namely that in these situations there is also a minimum mass solution, beyond which solutions can no longer be described by black holes \cite{Kanti:1995vq,Sotiriou:2014pfa}. In the shift-symmetric case, it was further noticed that an inner singularity and the horizon overlap in this limit \cite{Sotiriou:2014pfa}. In this section we explore this small mass limit of EsGB black holes for a generic coupling function, discussing the domains of existence of solutions.
To explore the small mass limit of static black holes in EsGB models we employ the static and spherically symmetric line element of Eq. \eqref{eq:lineelement}, for which the field equations are presented in Appendix \ref{app:feqs_static}. As already noted 
above no analytic solutions are known, and so numerical methods must be used.

\begin{figure}[]
\centering
\includegraphics[width=0.5\textwidth]{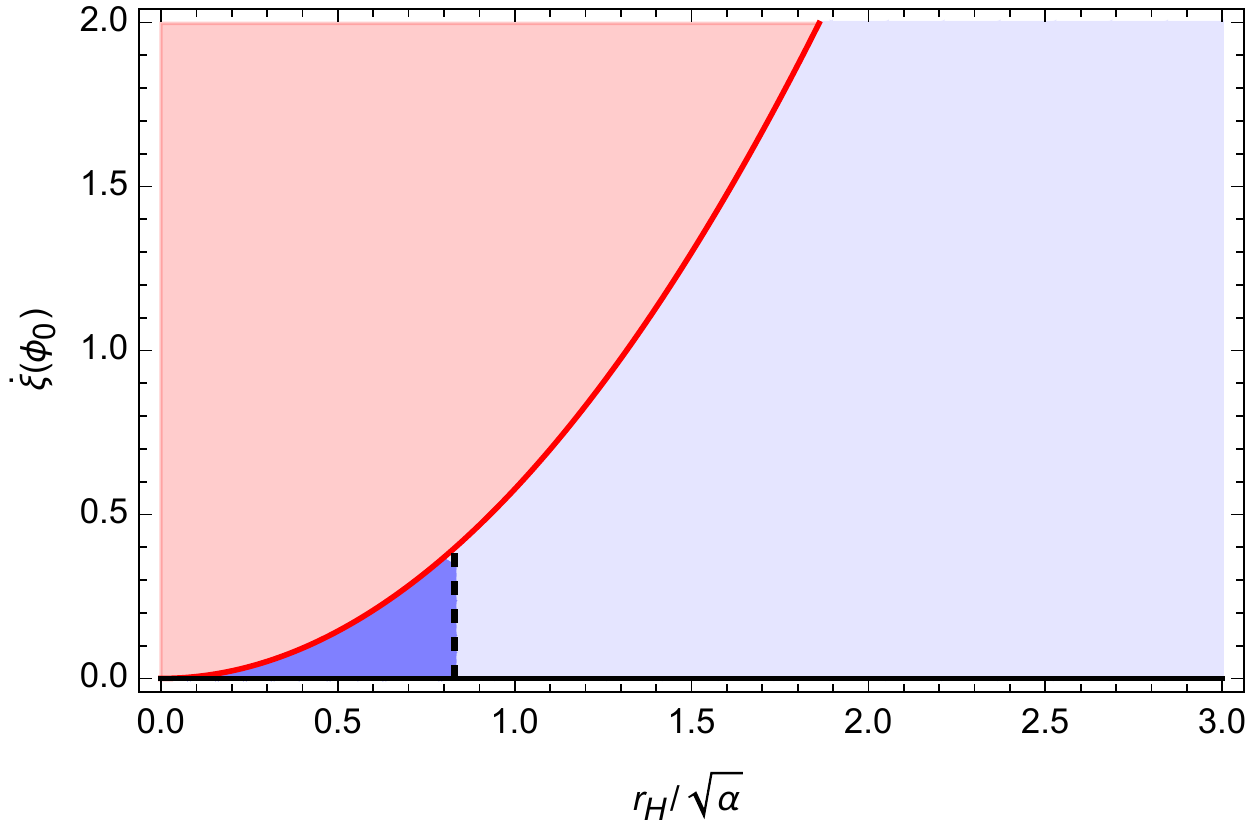}
\caption{Domain of existence of black hole solutions with scalar hair (blue region), obtained by plotting the condition of Eq. \eqref{eq:condition} for a general coupling. The darker blue region denotes the region where the Schwarzschild black holes are unstable (\textit{c.f.} Eq. \eqref{eq:condition2}), for couplings obeying the conditions of Eq. \eqref{eq:coupling_conditions}. Beyond the red line, solutions can no longer be described by black holes.}
\label{fig:domain-general}
\end{figure}

Nonetheless, near the boundaries of our domain, analytic methods can be employed. We assume that a static black hole solution of the model allows the asymptotic expansion near the event horizon (hereby denoted by $r_H$)
\begin{equation}
f(r) = \sum_{n=1}^{\infty} f_n \, \epsilon^n, \qquad
\delta(r) = \sum_{n=0}^{\infty} \delta_n \, \epsilon^n ,\qquad
\phi(r) = \sum_{n=0}^{\infty} \phi_n \, \epsilon^n, \qquad \epsilon \equiv \frac{r}{r_H}-1
\label{eq:horizon-expansion}
\end{equation}
then, the near-horizon field equations to zeroth order in $\epsilon$ become 
\begin{equation}
\begin{aligned}
&\mathcal{E}^{t}_{\phantom{t} t} = \mathcal{E}^{r}_{\phantom{r} r} = \frac{2r_H^2 - 2f_1 r_H^2 + \alpha f_1 \phi_1 \dot \xi(\phi_0)}{2r_H^4} = 0,\\&
\mathcal{E}^{\theta}_{\phantom{\theta} \theta} = \mathcal{E}^{\varphi}_{\phantom{\varphi} \varphi} = \frac{\left(f_1 (-2+3\delta_1) -2f_2\right)r_H^2 + \alpha f_1^2 \phi_1 \dot \xi(\phi_0)}{2 r_H^4} = 0,\\&
\mathcal{E}_\phi = \frac{2f_1 \phi_1 r_H^2 + \cbr{3\delta_1 f_1 +f_1^2 -2f_2}\alpha \dot \xi(\phi_0) }{2r_H^4} = 0,
\end{aligned}
\end{equation}
which can be solved for $f_1$ and $\phi_1$ in terms of $r_H$ and $\phi_0$, implying the following relations 
\begin{equation}
f_1 = \frac{2 r_H^2}{r_H^2 + \sqrt{r_H^4-3 \alpha^2 \dot \xi (\phi_0)^2}}, \qquad
\phi_1 = \frac{-r_H^2 + \sqrt{r_H^4 - 3\alpha^2 \dot \xi(\phi_0)^2}}{\alpha \dot \xi (\phi_0)}.
\end{equation}
From these expressions one immediately finds the remarkable result that the horizon radius has a minimum size beyond which the solution can no longer be described by a black hole\footnote{Hereafter we assume $\alpha>0$.}. This follows by
imposing the reality condition that $f_1$ and $\phi_1$ must be real, leaving us with the condition
\begin{equation}
r_H^4 - 3\alpha^2 \dot \xi(\phi_0)^2 \geq 0  \Leftrightarrow r_H \geq r_H^{min} = \sqrt{\sqrt{3} |\alpha \dot \xi(\phi_0)|}
\label{eq:condition}
\end{equation}
or equivalently,
\begin{equation}
    \qquad \dot \xi(\phi_0) \leq \frac{1}{\sqrt{3}}\left(\frac{r_H}{\sqrt{|\alpha|}}\right)^2\,.
\end{equation}
The above condition defines a region in the $(r_H/\sqrt{\alpha},\dot \xi(\phi_0))$ plane within which BH solutions with a regular (real) scalar field configuration can exist. This is illustrated in Fig. \ref{fig:domain-general}, where we have also highlighted the spontaneous scalarization condition \eqref{eq:condition2}. Note, however, that the minimum horizon radius $r_H^{min}$ depends on $\dot \xi (\phi_0)$. It is, therefore, possible that models where the minimum mass solution as given by Eq. \eqref{eq:condition} is never reached, depending on the behavior of $\phi_0$ and of the coupling function. Indeed, as mentioned in the introduction, in Ref. \cite{Doneva:2017bvd} the authors follow the fundamental scalarized branch all the way to vanishing masses.

\par To further explore the small mass limit of EsGB black holes, consider again the field equations given in Appendix \ref{app:feqs_static}, where we note that a closed-form expression for $\delta'(r)$ can be obtained by a simple algebraic manipulation of the $(r,r)$ equation. Substituting the value of $\delta'$ onto the other field equations we can further rewrite the whole system of field equations in matrix form\footnote{Here we have used the $(t,t)$ and the scalar field equations.}
\begin{equation}
    \mathbf{\mathcal{M}} \mathbf{x}' = \mathbf{b}
    \label{eq:system_mat}
\end{equation}
where $\mathbf{\mathcal{M}}$ and $\mathbf{b}$ are a $2\times 2$ matrix and a $2\times 1$ column vector respectively, whose components are given in Appendix \ref{app:feqs_static}, and $\mathbf{x} = \left[f(r)\,\, \phi'(r) \right]^T$. Given the above system of equations, and appropriate initial conditions at some point $r=r_0$, the existence theorem asserts that a solution to the Cauchy problem to extend our solution to a neighbouring point $r_1$ will exist if (see e.g. \cite{Alexeev:1996vs}, which also studies the singularity structure of dilatonic black holes)
\begin{equation}
    \mathrm{det} \mathbf{\mathcal{M}}|_{r=r0} \neq 0.
\end{equation}
Therefore if a point $r=r^*$ exists such that the determinant of $\mathbf{\mathcal{M}}$ vanishes, the system of field equations will be ill-posed at that point. The existence of such a point would indicate the presence of a coordinate or physical singularity, whose nature would have to be studied by other means. From a numerical point of view, any standard strategy used for numerical integration 
will stop before $r^*$. The determinant of $\mathbf{\mathcal{M}}$ in Eq. \eqref{eq:system_mat} is given explicitly by
\begin{equation}
    \mathrm{det} \mathbf{\mathcal{M}} = \frac{f\left(\alpha^3 f^3 \mathcal{A} + \alpha^2 f^2 \mathcal{B} + \alpha f \mathcal{C} + \mathcal{D} \right)}{2 r^4 \left(\alpha  (1-3 f) \phi' \dot \xi + 2 r\right)^2}
    \label{eq:det}
\end{equation}
where the expressions for $\mathcal{A},\mathcal{B},\mathcal{C},\mathcal{D}$ are again given in Appendix \ref{app:feqs_static}. A simple examination of the determinant \eqref{eq:det} reveals the existence of two zeros at the locations where $f(r)=0$, and where
\begin{equation}
    \alpha^3 f^3 \mathcal{A} + \alpha^2 f^2 \mathcal{B} + \alpha f \mathcal{C} + \mathcal{D} = 0.
    \label{eq:det_rs}
\end{equation}
The first, is related to the location of the event horizon ($r^*=r_H$) and the correspondent singularity is a coordinate one. The second case is more intricate, and is related to a curvature (physical) singularity, as we will see.

\par As a toy model to help us understand this singularity, consider Eq. \eqref{eq:det_rs} for a Schwarzschild background. This can be seen as the zeroth order solution in an expansion in $\alpha$ of the dilatonic and linear EsGB models.
The solution is 
\begin{equation}
    r^* = 6^{1/6} \left(M \alpha \right)^{1/3}.
\end{equation}
Similarly to the analytical example in Eq. \eqref{eq:4degb_sing}, we observe that, at least for very small couplings, there is a singularity obeying (approximately) a proportionality relation $r_s \propto \left(M \alpha \right)^{1/3}$. 

\par Let us now explore the behavior of Eq. \eqref{eq:det_rs} near the event horizon of a EsGB black hole. Using the same near-horizon expansion as before (Eq. \eqref{eq:horizon-expansion}), we observe that
\begin{equation}
    \begin{aligned}
        &\alpha^3 f^3 \mathcal{A} + \alpha^2 f^2 \mathcal{B} + \alpha f \mathcal{C} + \mathcal{D} = -4 r_H^5 \left(1-\left(r_H^{min}/r_H\right)^4+\sqrt{1-\left(r_H^{min}/r_H\right)^4} \right) + \mathcal{O}\left(\epsilon\right),
    \end{aligned}
    \label{eq:det_overlap}
\end{equation}
where $r_H^{min}$ was defined in Eq. \eqref{eq:condition}. Thus we observe that Eq. \eqref{eq:det_rs} vanishes at the event horizon if $r_H = r_H^{min}$, indicating the presence of a singularity (other than the typical coordinate one). Therefore, we conclude that in the limit $r_H \to r_H^{min}$, an overlap of the curvature singularity and the event horizon occurs.

\par In the following sections we will explore the small mass limit of EsGB black holes in more detail, utilising non-linear numerical solutions, and taking their inner structure into account. 

\subsection{Physical Quantities of Interest}
\par To numerically integrate the field equations in order to obtain the black hole solutions, we match the near-horizon expansion of Eq. \eqref{eq:horizon-expansion} with the appropriate asymptotic behaviour in  
far field ($r\to \infty$) limit:
\begin{equation}
f(r) =1 -\frac{2M}{r} + \mathcal{O}\left(r^{-2}\right)\,, \qquad \delta(r) = \mathcal{O}\left(r^{-2}\right),\qquad
\phi(r) = \frac{Q_s}{r} + \mathcal{O}\left(r^{-2}\right)\,.
\label{eq:asymptotics}
\end{equation}
where $M$ is the ADM mass and $Q_s$ the scalar charge of the solution.


\par Horizon quantities of physical interest include the Hawking temperature $T_H$, the horizon area $A_H$, and the entropy $S$. For the line element \eqref{eq:lineelement} these are given by
\begin{equation}
    T_H = \frac{1}{4\pi} f_1 e^{-\delta_0}, \qquad A_H = 4\pi r_H^2,
\end{equation}

\begin{equation}
    S = \frac{1}{4}A_H + \frac{\alpha}{8}\int_H d^2x \sqrt{h} \xi \cbr{\phi} R^{(2)} = \frac{1}{4}A_H + \pi \alpha \xi\cbr{\phi_0}
\end{equation}
where $h$ is the determinant of the induced metric on the horizon, and $R^{(2)}$ its Ricci scalar \cite{Iyer:1994ys}. The horizon and asymptotic quantities are related by a Smarr-type relation given by \cite{Iyer:1994ys,Liberati:2015xcp}
\begin{equation}
    M + M_s = 2T_H S,
    \label{eq:smarr_static}
\end{equation}
where
\begin{equation}
    M_s = -\frac{1}{4\pi} \int d^3x \sqrt{-g} \frac{\xi(\phi)}{\dot \xi(\phi)} \Box \phi.
\end{equation}
For the dilatonic coupling \eqref{eq:coupling-exp} the above relation simplifies to $M_s = Q_s/\gamma$. Furthermore, it can be shown that for the linear coupling \eqref{eq:coupling-linear} the following condition holds \cite{Prabhu:2018aun}
\begin{equation}
    Q_s = 2\pi \alpha T_H.
    \label{eq:linear_relation}
\end{equation}
These conditions can be used to estimate the accuracy of our numerical method.

\par Once again, an interesting remark can be made about the Hawking temperature, that in the small mass/size limit gives
\begin{equation}
    \lim_{r_H \to r_H^{min}} T_H = \frac{e^{-\delta_0}}{2\pi r_H^{min}} > 0\,
\end{equation}
and thus, as in the case of section \ref{sec:4DEGB}, evaporation will not halt in the small mass limit and the black hole will continue to lose its mass at a rate given by Eq. \eqref{eq:greybody}, posing a potential threat to cosmic censorship\footnote{See however Ref. \cite{Ong:2019glf} (and Ref. \cite{Ong:2020xwv} for a review) where a similar situation occurs for charged dilatonic black holes in the Einstein-Maxwell-scalar theory, but where cosmic censorship holds nonetheless.}.

\subsection{Numerical Method}
We now compute numerical solutions to the field equations. We use a Runge-Kutta-45 ordinary differential equation solver and implement a shooting method for the parameter $\phi_0$ such that the asymptotic expansions are matched. In more detail, the near horizon expansion of Eq.~\eqref{eq:horizon-expansion} is 
used to set initial conditions for a numerical integration, with the only free parameter being $\phi_0$ (once $r_H$ and $\alpha$ are fixed). The field equations are 
then integrated from the horizon outwards to large $r$, the result is compared with the asymptotic expansion at 
large $r$, $\phi_0$ adjusted, and the procedure 
repeated until the results of the numerical 
integration match the asymptotic expansion. Finally, using the results for the shooting parameters, the field equations are integrated from the horizon inwards to probe the internal structure of the black hole. We monitor curvature scalars such as the Ricci and GB scalars throughout the domain of integration, along with the determinant presented in Eq. \eqref{eq:det}. To test the accuracy of the numerical solutions we use the relations \eqref{eq:smarr_static} and \eqref{eq:linear_relation}. We remark that errors are on the order of $10^{-8}$.

\subsection{Numerical Results}
\par Using the numerical algorithm described, we have explored the linear \eqref{eq:coupling-linear}, dilatonic \eqref{eq:coupling-exp} and quadratic exponential \eqref{eq:coupling-quad} couplings. For the latter, we explore several values of $\beta$.

\subsubsection{Linear and Dilatonic Couplings}

\begin{figure}[]
\centering
\includegraphics[width=0.5\textwidth]{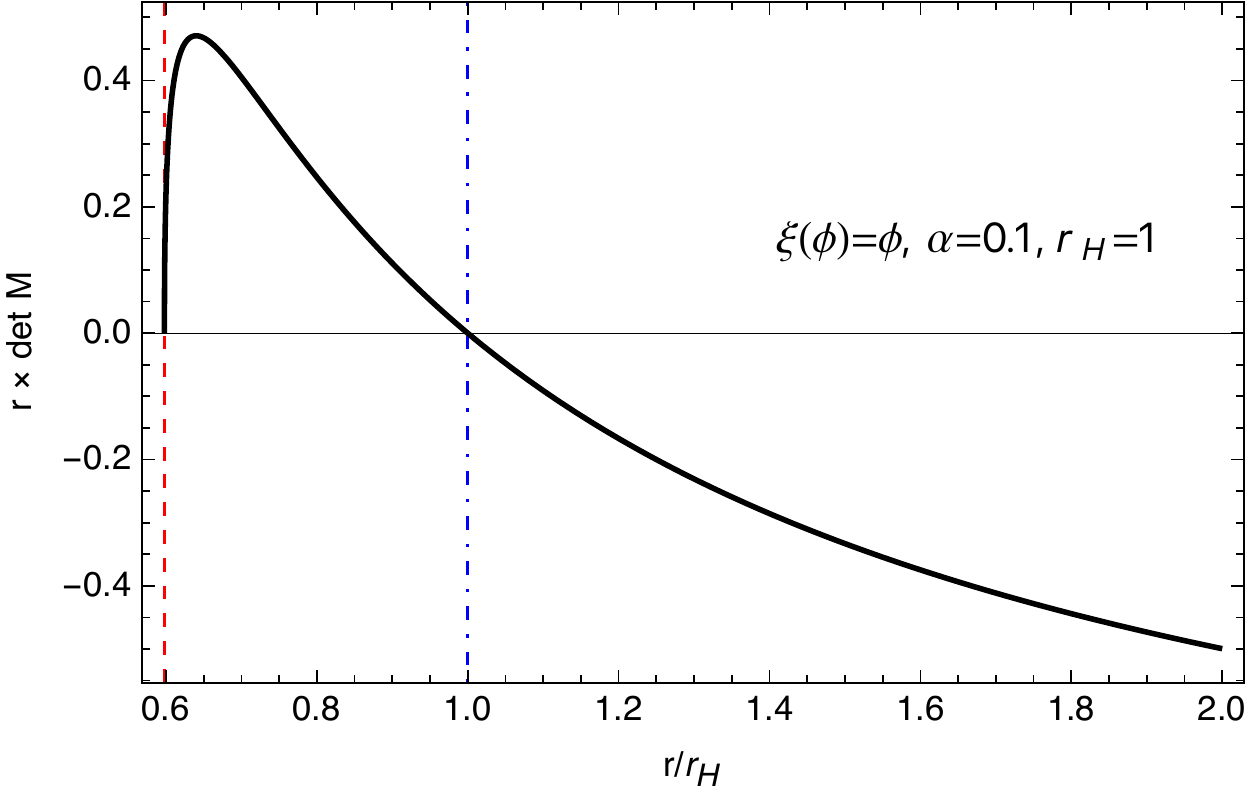}
\caption{Determinant presented in Eq. \eqref{eq:det} for the solution of the plot on the left in Fig. \ref{fig:metricfunctions}. We observe that $\mathrm{det} \mathbf{\mathcal{M}}$ has two zeros, one at the event horizon (blue dot-dashed line) and another at the singularity $r_s$ (red dashed line).}
\label{fig:determinant_1}
\end{figure}

For the linear and dilatonic couplings, by monitoring the Ricci and GB scalars, a finite radius singularity was always found at a radius $r=r_s>0$ inside the horizon, whose value ultimately depends on the ratio between the horizon radius and the coupling $\alpha$. The singularity is located where $\mathrm{det} \mathbf{\mathcal{M}}$ in Eq. \eqref{eq:det} vanishes inside the event horizon, as observed in Fig. \ref{fig:determinant_1}. 

\begin{figure}[]
\centering
\includegraphics[width=0.5\textwidth]{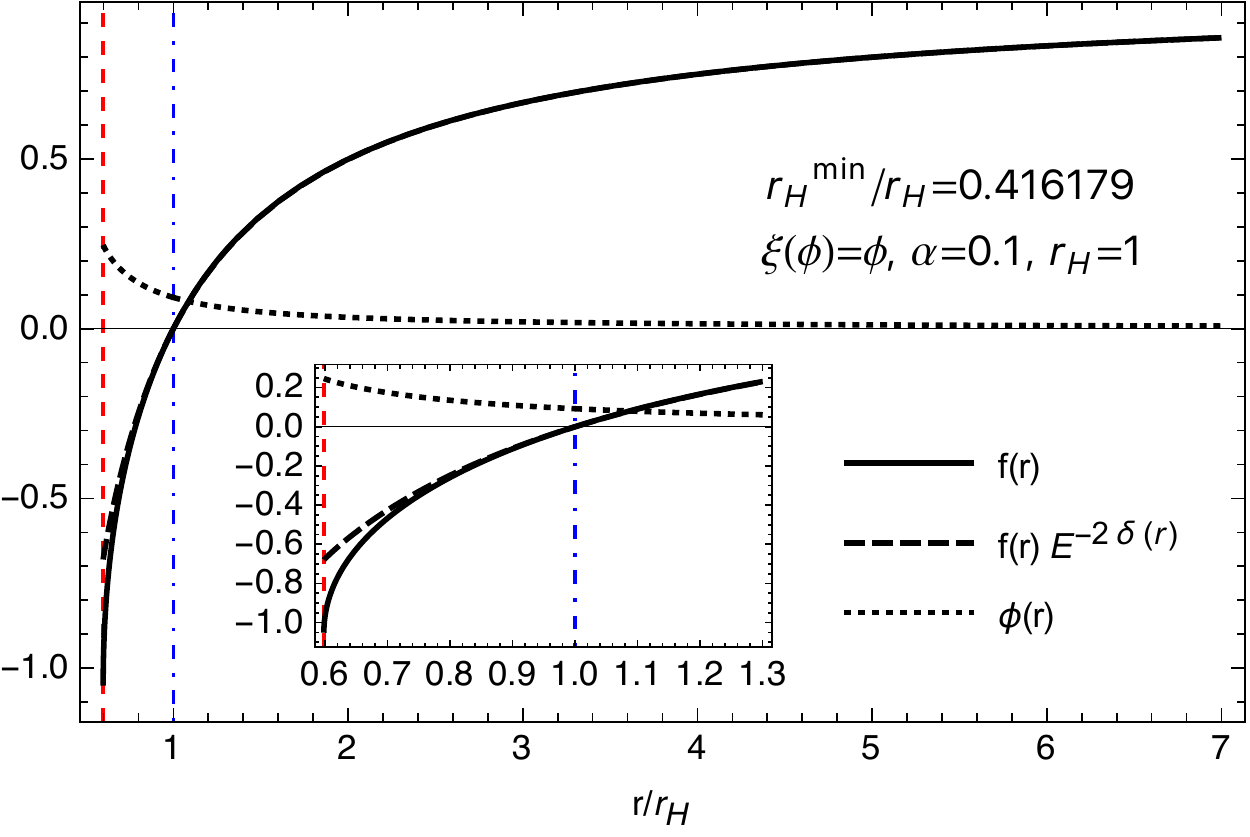}\hfill
\includegraphics[width=0.5\textwidth]{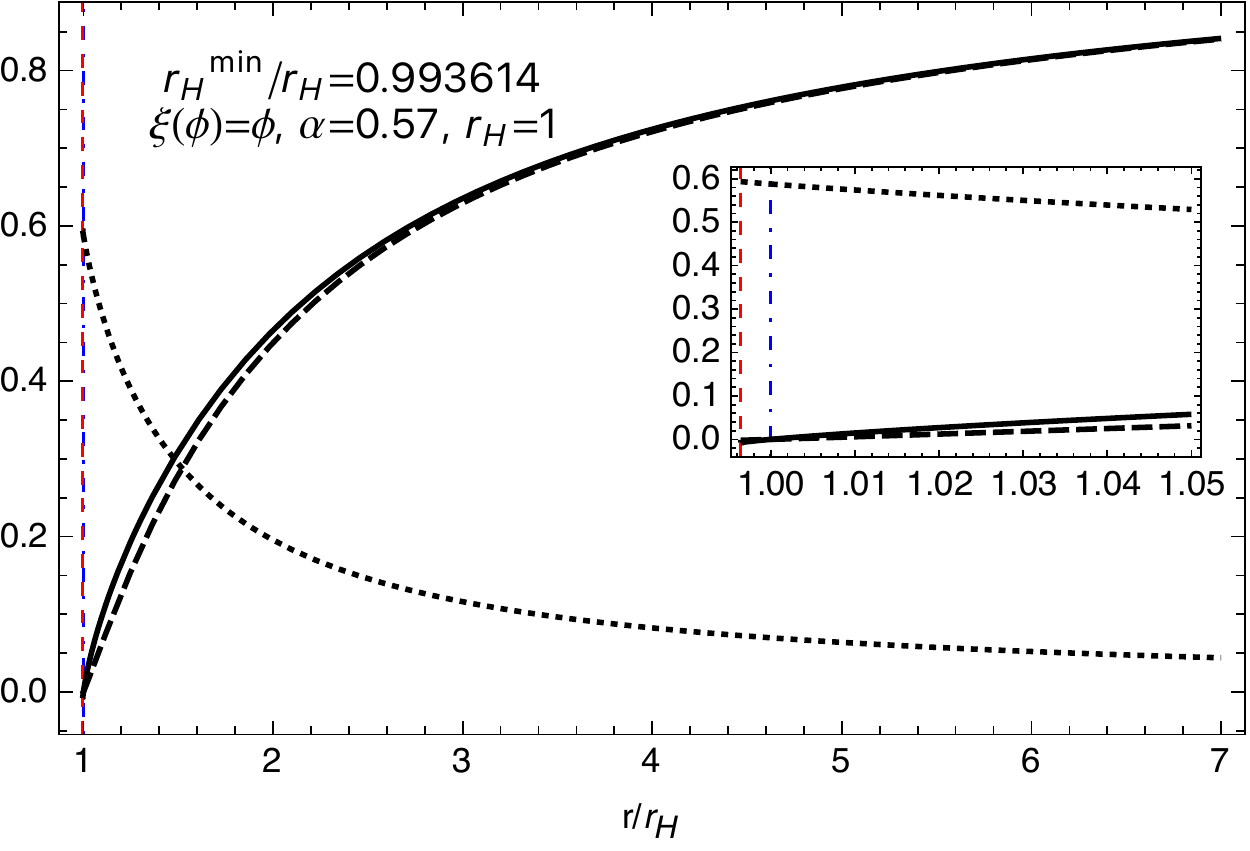}\hfill
\caption{Metric functions and scalar field for two different values of $r_H/\sqrt{\alpha}$. The blue vertical line denotes the event horizon, while the red one denotes the location of the finite radius singularity $r_s$.}
\label{fig:metricfunctions}
\end{figure}

\begin{figure}[]
\centering
\includegraphics[width=0.5\textwidth]{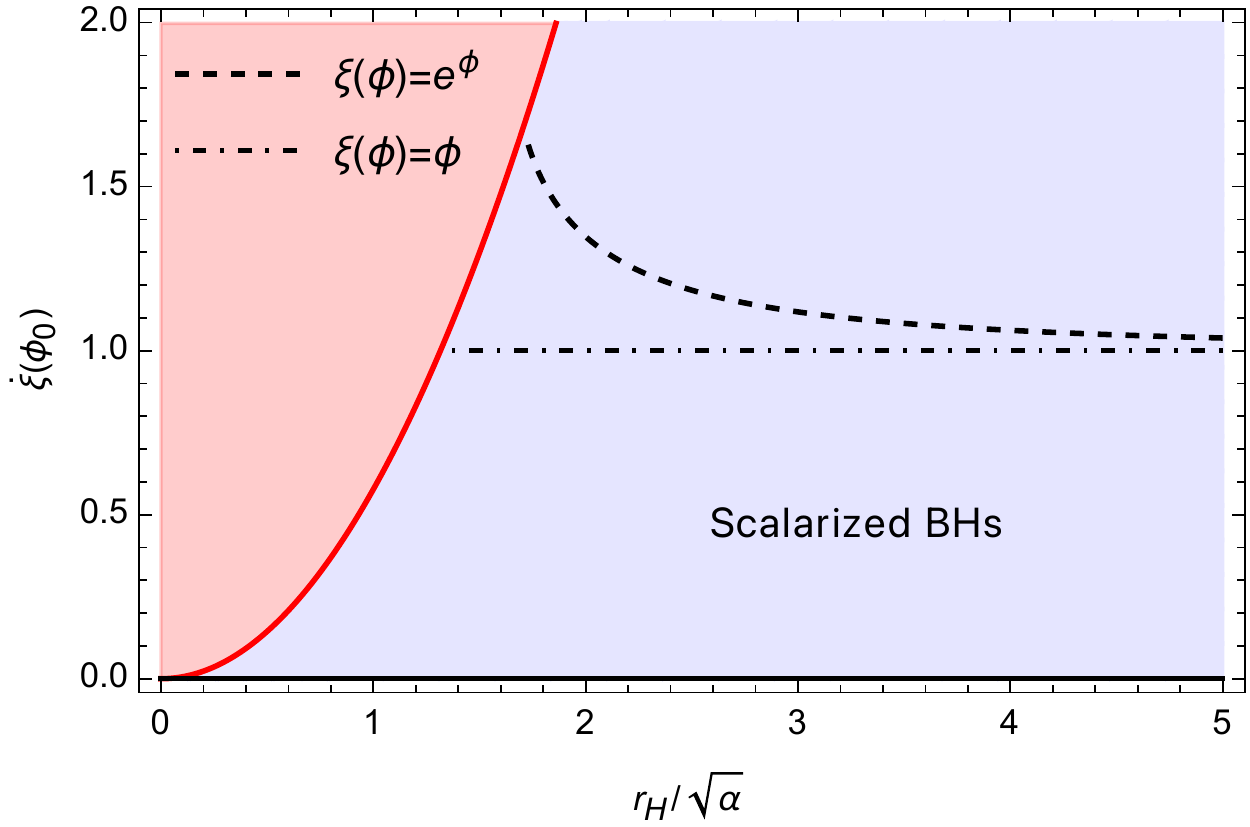}\hfill
\includegraphics[width=0.5\textwidth]{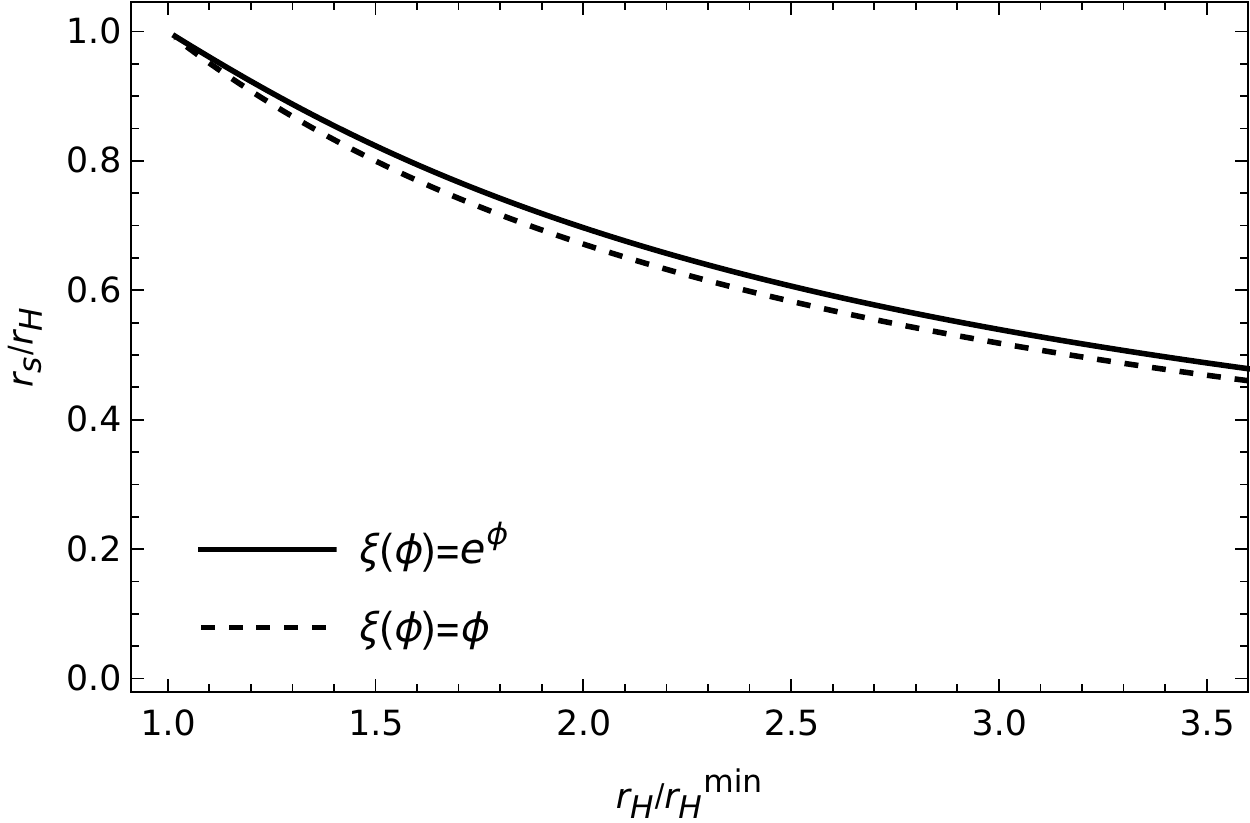}
\caption{On the left we observe the domain of existence of black holes for the exponential (dashed line) and linear (dot-dashed line) couplings, delimited by the singular line in red. The intersection of the red line with the black ones denote the point in the domain where the finite radius singularity and the event horizon overlap, as $r_H \to r_H^{min}$, as observed in the figure on the right, for both couplings.}
\label{fig:domain-lin-exp}
\end{figure}

In Fig. \ref{fig:metricfunctions}, we plot the metric functions along with the locations of $r_s$ and $r_H$ for several values of $r_H^{min}/r_H$.
As the horizon radius approaches $r_H^{min}$ (as one would expect to happen dynamically as  Hawking evaporation proceeds), the horizon and the singularity overlap, and numerical solutions reveal divergences of the curvature invariants, derivatives of the metric functions and the scalar field. That the location of $r_s$ and $r_H$ overlap when $r_H \to r_H^{min}$ is in agreement with our analytical exploration shown in Eq. \eqref{eq:det_overlap}, and is similar to the behaviour observed in the analytical example of Section \ref{sec:4DEGB} (c.f. Fig. \ref{fig:rsrh4degb} and Fig. \ref{fig:domain-lin-exp}, right).

The domains of existence were constructed for both couplings, and can be observed in Fig. \ref{fig:domain-lin-exp} (left), where each point on the dashed and dot-dashed lines represent a numerical black hole solution. Note that the domains of existence end at the red line, where $r_H=r_H^{min}$. Assuming that Hawking radiation gradually reduces the mass of a black hole (and hence $r_H$) for some fixed $\alpha$, but that our numerical solutions instantaneously remain an accurate approximation, we see that the fate of all black holes for both these couplings is to follow the lines on  Fig.~\ref{fig:domain-lin-exp} and to always to reach the red line. 

\subsubsection{Quadratic exponential (spontaneous scalarization) coupling}
\begin{figure*}[t!]
\centering
\includegraphics[width=0.5\textwidth]{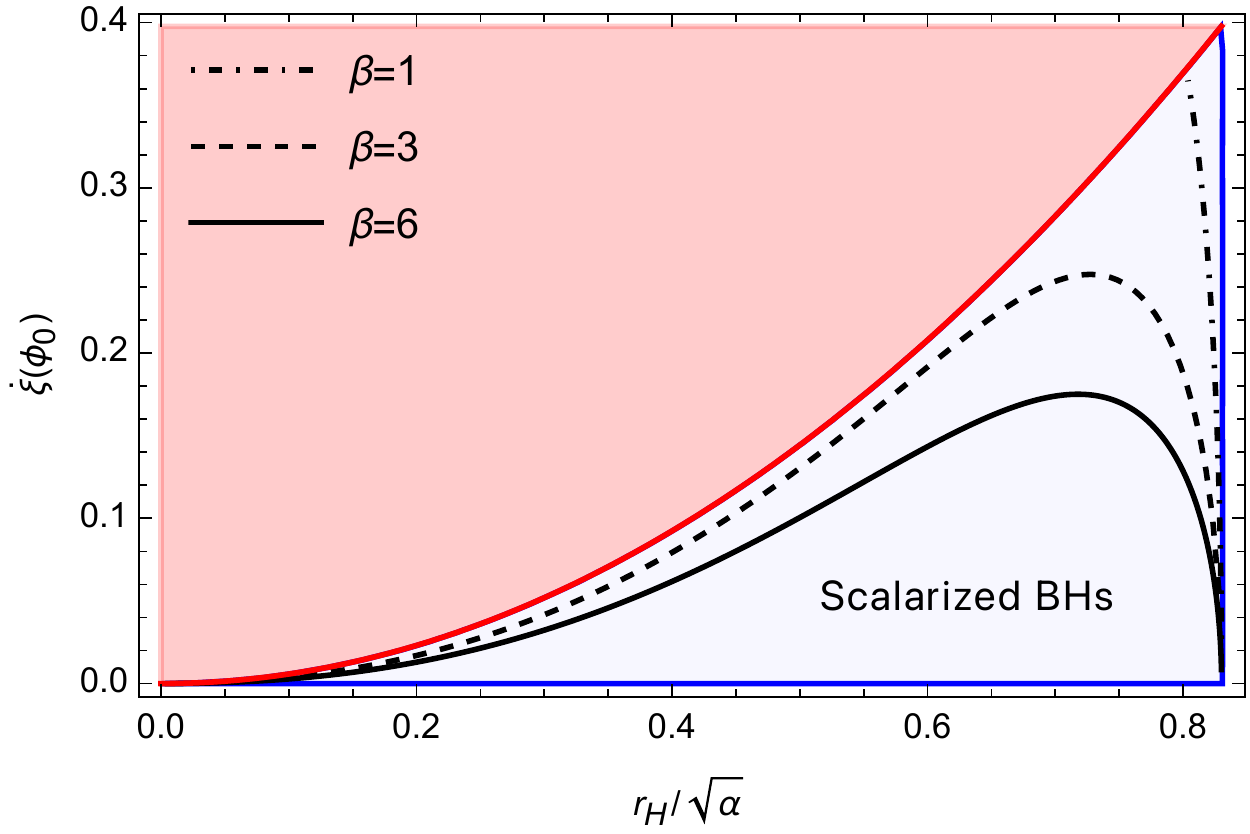}\hfill
\includegraphics[width=0.5\textwidth]{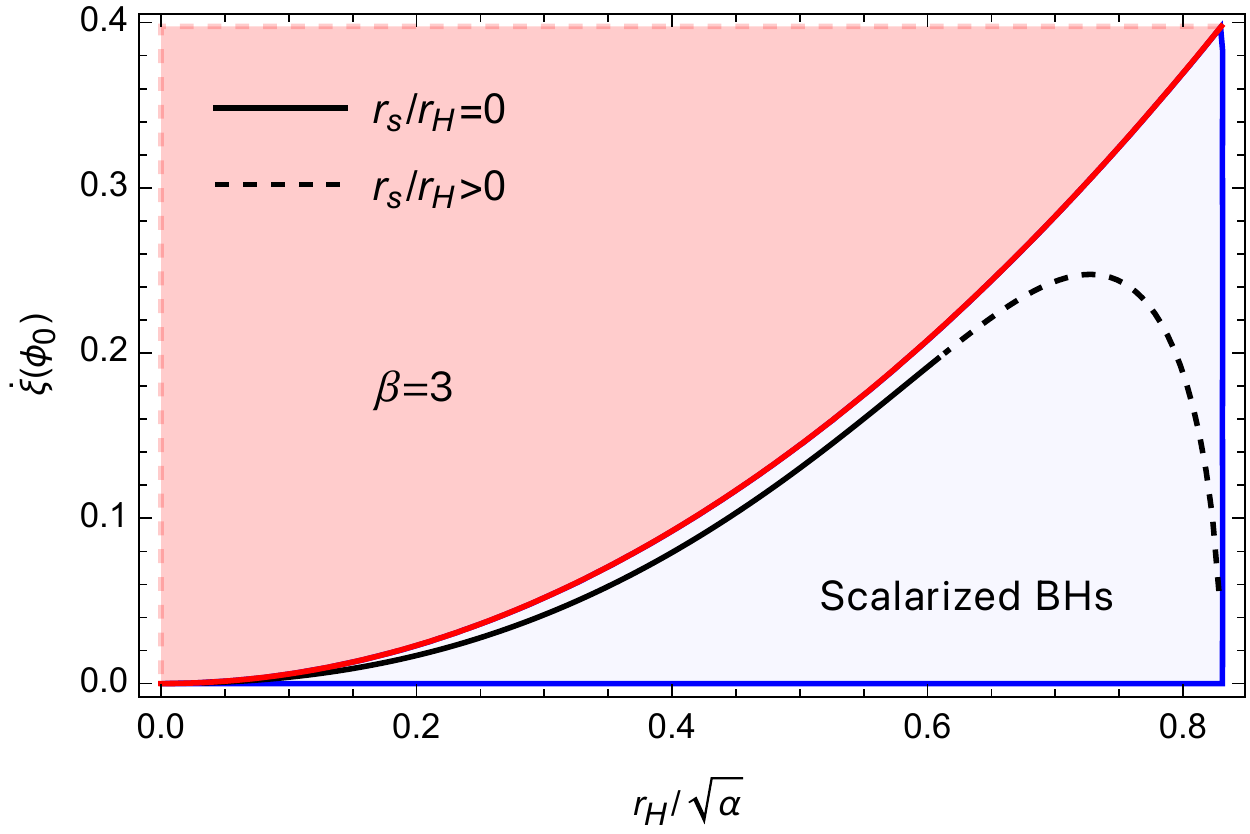}\vfill
\includegraphics[width=0.5\textwidth]{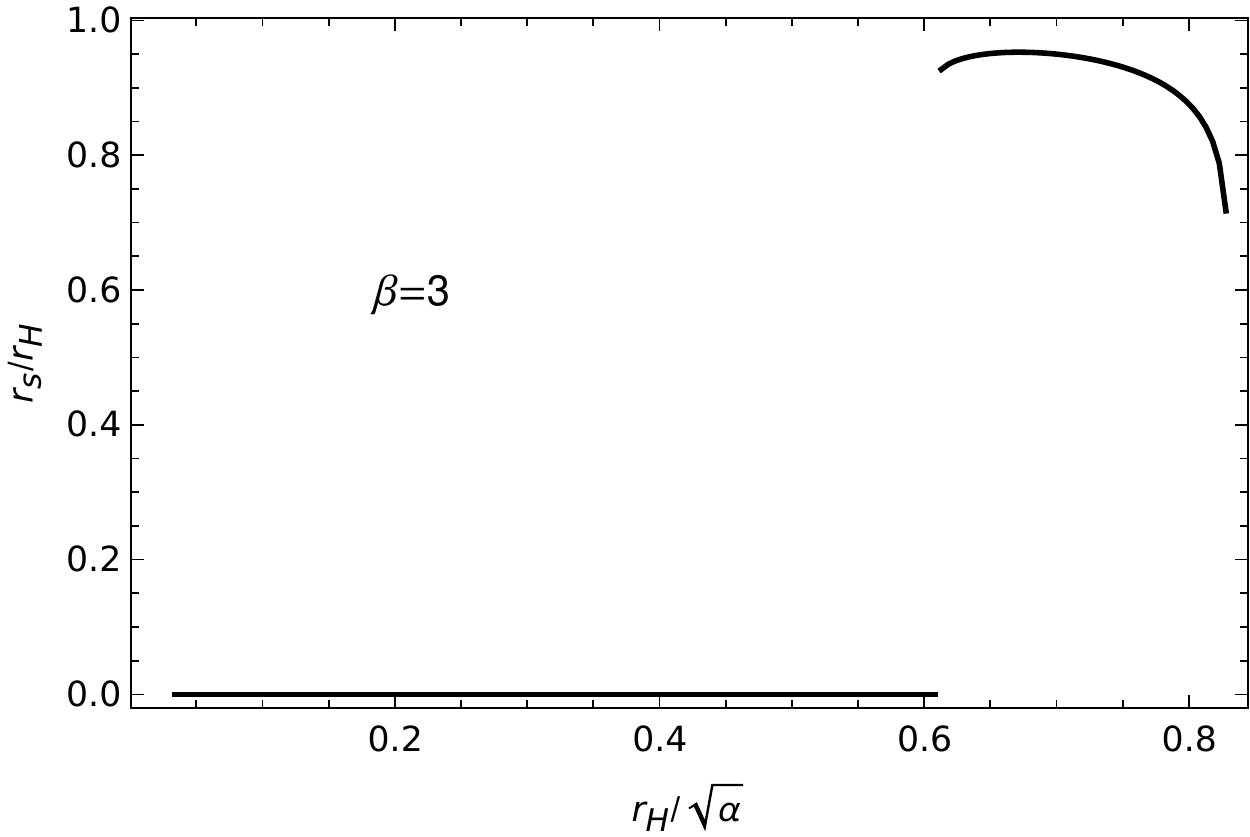}\hfill
\includegraphics[width=0.5\textwidth]{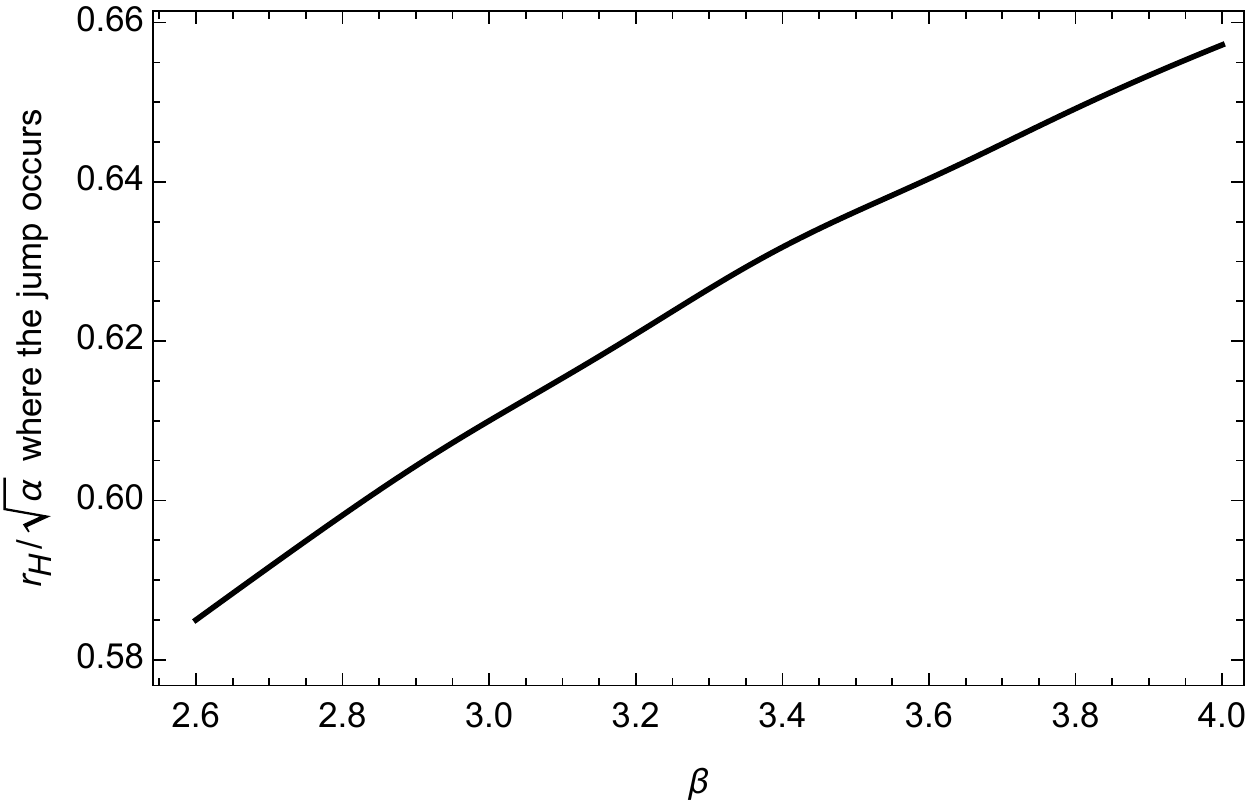}
\caption{(Top Left) Domain of existence of spontaneously scalarized solutions for $\beta=1,3,6$. For values $\beta < \beta_{\mathrm{crit}}$, the characteristics of the domain of existence are similar to those of the dilatonic and linear couplings, where the inner singularity and the horizon overlap as the BH shrinks. For larger values of $\beta$, solutions never reach $r_H=r_H^{min}$, and can shrink all the way down to $r_H=0$. (Top Right) Behavior of the inner finite radius singularity for $\beta=3$. A finite radius singularity with $r_s >0$ exists only until the black hole shrinks to a certain $r_H/\sqrt{\alpha}$ value, beyond which there is no singularity other than at the origin. This can be observed at the bottom figure (left). The location where the jump in the singularity behaviour occurs is presented in the bottom right, for a range of $\beta$.}
\label{fig:domain-quad}
\end{figure*}

Consider now the coupling of Eq. \eqref{eq:coupling-quad}. We will perform a similar analysis as before, for several values of $\beta$. Note that higher values of $\beta$ suppress scalarization. The domains of existence for $\beta=1,3,6$ can be observed in Fig. \ref{fig:domain-quad} on the top left. We observe that for the $\beta=1$ case, the domain of existence of solutions is similar to the dilatonic and linear coupling cases, where the location of the horizon and singularity overlap as the black hole shrinks, terminating in a critical solution. However, for $\beta=3$ and $6$, the domain of existence is radically different from the previous cases, as solutions never reach the singular (red) line, allowing the black hole to shrink to $r_H \sim 0$. This is possible due to the dependence of $r_H^{min}$ on the value of the derivative of the coupling function at the horizon (as in Eq. \eqref{eq:condition}), and this behavior was observed for values
\begin{equation}
    \beta > \beta_{\mathrm{crit}} \approx 2.33125,
    \label{eq:betacrit}
\end{equation}
while for lower values the black line would intersect the red one, the domain of existence ending in a critical solution.

The behaviour of the inner finite radius singularity is rather curious for $\beta > \beta_{\mathrm{crit}}$. We find there exists a finite radius singularity with $r_s >0$ only until to a certain value of $r_H/\sqrt{\alpha}$, below which there is no singularity other than at the origin. This is shown in Fig. \ref{fig:domain-quad} (top right) for $\beta=3$, where the dashed line show the part of the domain of solutions where $r_s>0$ and the solid line the part for which $r_s=0$. The transition is abrupt as shown in Fig. \ref{fig:domain-quad} (bottom left), where the location of $r_s$ is plotted as a function of $r_H$. On the bottom right in Fig. \ref{fig:domain-quad} we plot the location of the jump for a range of $\beta$. We have performed simulations for very large $\beta$ of $\mathcal{O}\left(10^2\right)$, and observed that the finite radius singularity exists only in a narrow of the domain of existence. For example, for $\beta=100$ the finite radius singularity exists only from $r_H/\sqrt{\alpha} \approx 0.83$ (as in Eq. \eqref{eq:condition2}) down to $r_H/\sqrt{\alpha} \approx 0.828$, and the maximum value of $r_s/r_H$ is about $0.5$.

\par
\begin{figure}[]
\centering
\includegraphics[width=0.5\textwidth]{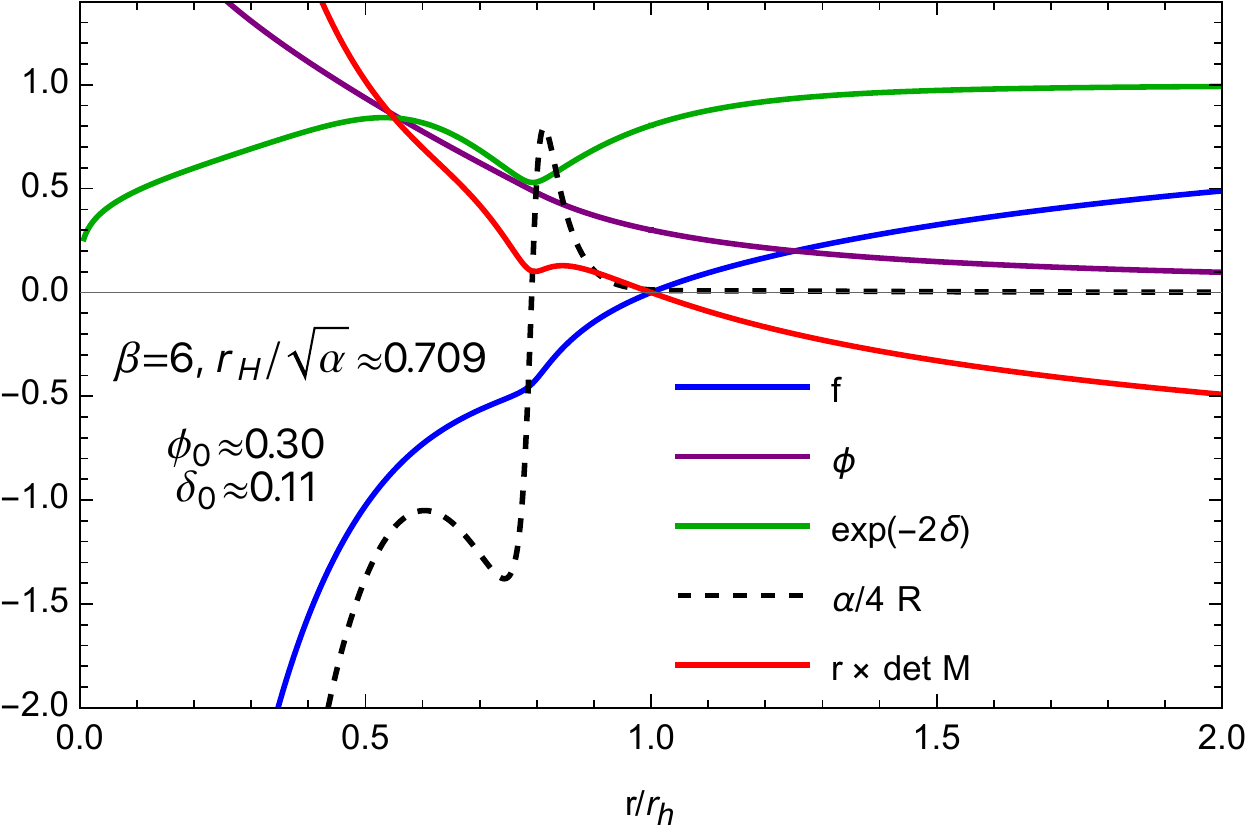}\hfill
\includegraphics[width=0.5\textwidth]{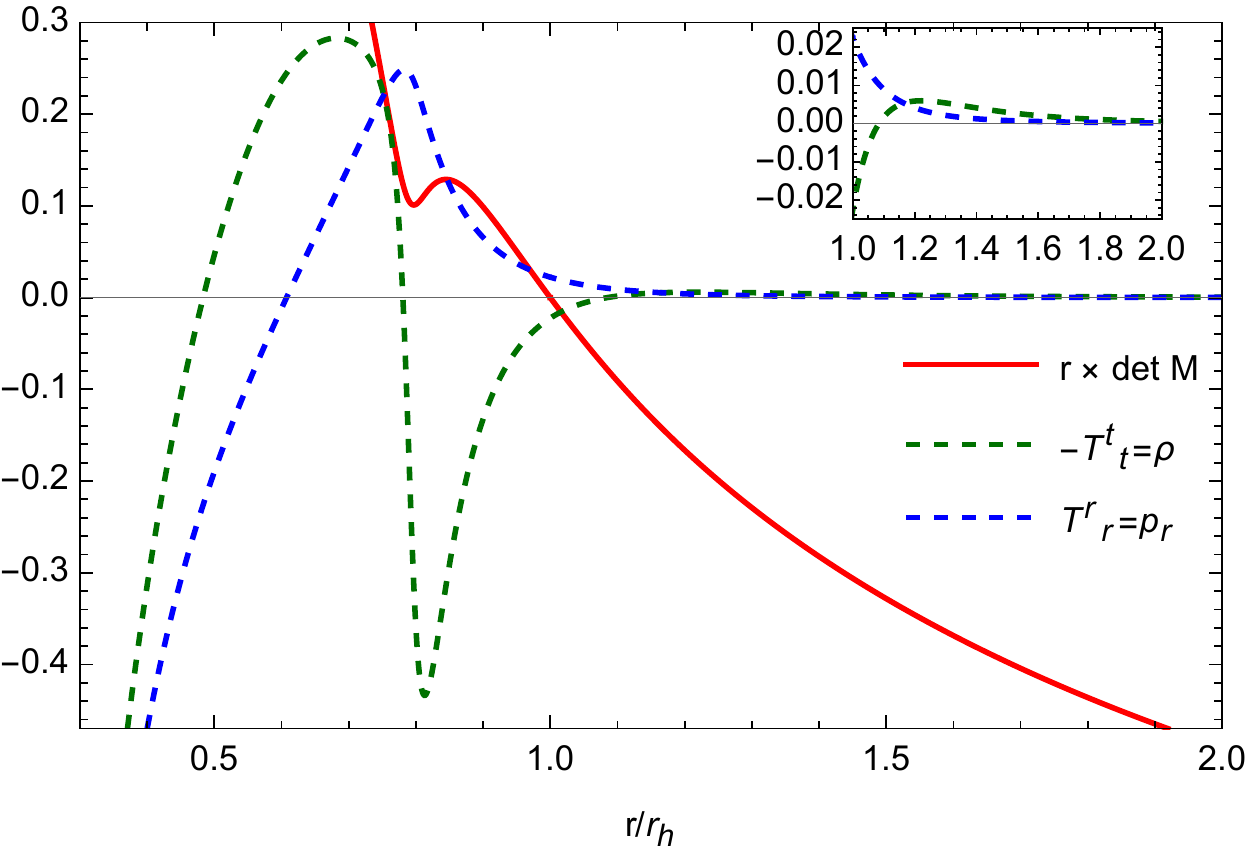}
\caption{
On the left we observe a fiducial black hole solution with no finite radius singularity (other than at $r=0$). Note that $\mathrm{det} \mathbf{\mathcal{M}}$ never vanishes inside the event horizon and the Ricci scalar is well-behaved all through the domain of integration (except at $r=0$). On the right we present the components of the stress-energy tensor $\rho$ and $p_r$ (scaled by a factor of $10^{-1}$ for presentation purposes) for the same solution, where we observe that repulsive effects are maximum near the turning point of the determinant.}
\label{fig:singularity-transition}
\end{figure}

 On Fig. \ref{fig:singularity-transition} we observe a fiducial numerical black hole solution (along with other relevant quantities) for which there is no finite radius singularity. Note that $\mathrm{det} \mathbf{\mathcal{M}}$ is strictly positive inside the event horizon. From a physical point of view we note that, as seen by the profile of the radial pressure $p_r$ in Fig. \ref{fig:singularity-transition} (right), repulsive effects are maximum near the turning point of $\mathrm{det} \mathbf{\mathcal{M}}$. The determinant then gets further away from zero as the repulsive effects get gradually weaker further inside the horizon.

\par An intuitive view on Hawking evaporation for this coupling would be the following. Starting from a (sufficiently large) Schwarzschild black hole, Hawking radiation will gradually reduce the mass of the black hole (for some fixed $\alpha$), until the condition of Eq. \eqref{eq:condition2} is met. A tachyonic instability would then settle in, leading to dynamical scalarization of the Schwarzschild black hole. The new scalarized solution will itself evaporate, with the endpoint now depending on the value of $\beta$. If $\beta < \beta_{\mathrm{crit}}$, the picture would be not too different to that of the dilatonic and linear couplings explored in the previous sections and Refs. \cite{Corelli:2022pio,Corelli:2022phw}, where the formation of a naked singularity is expected. However, if $\beta > \beta_{\mathrm{crit}}$, the evaporation process is expected to be similar to that of a Schwarzschild black hole, going all the way down to scales where quantum effects are expected to be important on general grounds, and our theory breaks down.

\par Remarkably, the behaviour of the inner singularity for the quadratic-exponential coupling shows that the critical solution end point of the domain of existence is \textit{not} a generic feature of gravitational theories with higher-order curvature terms.

\section{Upper bounds on the coupling constant}
\label{sec:constraints}
\par 
If evaporation proceeds as expected the behaviour we have been illustrating indicates that the small mass regime of EsGB theories may constrain the {\it allowed form of couplings} through self consistency arguments.
However, even if evaporation is not taken into account, the minimum allowed size of black holes can place constraints on the strength of the allowed coupling through observational constraints.

To do so it is important to take a different form of the action such that results are 
consistent across the literature, and thus imposed on equivalent definitions of the coupling constant. When discussing observational constraints, the action for EsGB theories is usually presented in the following form \cite{Lyu:2022gdr,Yunes:2016jcc,Perkins:2021mhb,Nair:2019iur,Yagi:2011xp}
\begin{equation}
\begin{aligned}
    \overline{S}=\int d^4x \sqrt{-g} \cbr{\frac{R}{16\pi}-\frac{1}{2}\left(\nabla \varphi \right)^2 + \overline{\alpha} \,F\cbr{\varphi} \GB}
    \label{eq:action_constraints}
\end{aligned}
\end{equation}
The mapping of the action \eqref{eq:action} to the above parametrization can be done as
\begin{equation}
    \phi = \sqrt{8\pi} \varphi, \qquad \alpha = 64\pi \overline{\alpha},
\end{equation}
and choosing $\xi(\phi)$ accordingly such that it is compatible with the definition of $F(\varphi)$. To be consistent with the literature, we will constraint the coupling constant $\overline{\alpha}$.

For each of coupling functions, we consider the singular static solution with $r_H=r_H^{min}$. Each of these singular black holes will have an associated minimum mass $M_{min}$. From our numerical black hole solutions we can extract the quantity
\begin{equation*}
    m_{min} \equiv M_{min}/\sqrt{\overline{\alpha}}.
\end{equation*}
Assume now that an observation was made, in which a black hole was measured to have mass $M_{obs}$. To be consistent with the description of a black hole within the EsGB theory we impose that the observed mass is greater than the allowed minimum mass
\begin{equation}
    M_{obs}/\sqrt{\overline{\alpha}} > m_{min}.
\end{equation}
Therefore, reintroducing the physical constants, from the above relation we obtain the bound
\begin{equation}
    \sqrt{\overline{\alpha}} < \frac{G M_{obs}}{c^2 m_{min}}.
    \label{eq:constraints-1}
\end{equation}
This last equation allows us to impose an upper bound on the coupling constant for each coupling function. From our numerical solutions we have extracted
\begin{equation}
\begin{aligned}
    &m_{min} \approx 4.66717, \qquad \mathrm{for} \qquad F\cbr{\varphi}=\varphi,\\&
    m_{min} \approx 4.91642, \qquad \mathrm{for} \qquad F\cbr{\varphi}=e^\varphi.
\end{aligned}
\end{equation}
Considering the case of GW190814 \cite{LIGOScientific:2020zkf}, where a compact object with a mass of around $M_{obs}=2.6 M_{\odot}$ was observed, and assuming it is a black hole, our calculations using Eq. \eqref{eq:constraints-1} give the upper bound
\begin{equation}
\begin{aligned}
    &\sqrt{\overline{\alpha}} \lesssim 0.82 \,\mathrm{km}, \qquad \mathrm{for} \qquad F\cbr{\varphi}=\varphi,\\&
    \sqrt{\overline{\alpha}} \lesssim 0.78 \,\mathrm{km}, \qquad \mathrm{for} \qquad F\cbr{\varphi}=e^\varphi.
\end{aligned}
\end{equation}
To the best of our knowledge, these constraints are the tightest constraints on $\overline{\alpha}$ so far, with the previous strongest upper bound being $\sqrt{\overline{\alpha}} \lesssim 1.18 \,\mathrm{km}$ for the linear coupling \cite{Lyu:2022gdr}. Constraints on the coupling obtained using data from other events can be found in Table \ref{tab:constraints}.

\begin{table}[h!]
    \begin{tabular}{|c|c|c|c|}
      \cline{3-4}
      \multicolumn{2}{}{} & \multicolumn{2}{|c|}{Upper bound on $\sqrt{\overline{\alpha}}$ (km)} \\ \hline
        Event/Ref. & $M_{obs}$ ($M_\odot$) & $F\cbr{\varphi}=\varphi$ & $F\cbr{\varphi}=e^\varphi$ \\\hline
      \hline
      GW190814 \cite{LIGOScientific:2020zkf} & $2.59 \pm 0.09$ & $0.82 \pm 0.03$ & $0.78\pm 0.03$\\
      \hline
      \cite{Jayasinghe:2021uqb} & $3.04\pm 0.06$ & $0.95 \pm 0.02$ & $0.91\pm 0.02$ \\
      \hline
      \cite{2019Sci...366..637T} & $3.30^{+ 2.8}_{-0.7}$ & $1.04^{+ 0.89}_{-0.22}$ & $0.99^{+ 0.84}_{-0.21}$ \\
      \hline
      GW200115 \cite{LIGOScientific:2021qlt} & $5.70^{+ 1.8}_{-2.1}$ & $1.80^{+ 0.57}_{-0.66}$ & $1.71^{+ 0.54}_{-0.63}$ \\
      \hline
    \end{tabular}
    \caption{Upper bounds on the coupling $\sqrt{\overline{\alpha}}$ obtained using data from several different events.}
    \label{tab:constraints}
\end{table}

We have not studied constraints on the quadratic-exponential coupling in Eq. \eqref{eq:coupling-quad} for several reasons. First, there is an important dependence on $\beta$, i.e., if $\beta > \beta_{\mathrm{crit}}$, then as discussed above, no minimum mass solutions exist and no upper bound can be imposed. Also, in the small $\beta$ limit we know solutions to be unstable \cite{Blazquez-Salcedo:2018jnn,Silva:2018qhn}. Secondly, there is no guarantee that the black hole in question is a scalarized black hole.

Note that our results should be taken with a pinch of salt given that EsGB models are not UV-complete. It is possible that the pathological behaviour in the small mass limit is cured by higher-order corrections to the theory, depending on the scale they at which they become relevant.

\section{Spinning Black Hole Solutions}
\label{sec:spin}
So far we have only studied static black hole solutions. In this section we will go one step further and explore spinning black hole solutions in EsGB theories. For this we resort, once again, to numerical integration of the field equations given in Section \eqref{sec:shape_coupling}. The  numerical procedure we adopt follows closely that of Ref. \cite{Delgado:2020rev}, Section 4. Namely, we consider a stationary and axi-symmetric line element of the form
\begin{equation}
    ds^2 = - e^{2\mathcal{F}_0}\left(1-\frac{\rho_H}{\rho}\right) dt^2 + e^{2\mathcal{F}_1} \left(\frac{d\rho^2}{1-\frac{\rho_H}{\rho}} + \rho^2 d\theta^2 \right) + e^{2\mathcal{F}_2}\rho^2 \sin^2 \theta \left(d\varphi-W dt\right)^2,
\end{equation}
where the functions $\mathcal{F}_i$, $W$ and the scalar field $\phi$ depend only on $\rho$ and $\theta$, and $\rho_H$ is the location of the event horizon in this coordinate system. The boundary conditions are implemented as follows. Asymptotic flatness is guaranteed by imposing
\begin{equation}
    \lim_{\rho \to \infty} \mathcal{F}_i = \lim_{\rho \to \infty} W = \lim_{\rho \to \infty} \phi = 0,
\end{equation}
while at the horizon, with the introduction of a new radial coordinate $x = \sqrt{\rho^2-\rho_H^2}$, we have
\begin{equation}
    \partial_x \mathcal{F}_i = \partial_x \phi = 0, \qquad W = \Omega_H, \qquad \mathrm{at} \qquad x = 0,
\end{equation}
where $\Omega_H$ is the horizon angular velocity. Axial symmetry and regularity impose
\begin{equation}
    \partial_\theta \mathcal{F}_i = \partial_\theta W = \partial_\theta \phi = 0,  \qquad \mathrm{at} \qquad \theta = 0, \pi.
\end{equation}
Focusing on black holes with parity reflection symmetry, we consider only the range $0 \leq \theta \leq \pi/2$, and impose the previous boundary conditions on the equator $\theta=\pi/2$ instead of $\theta=\pi$. The absence of conical singularities further imposes that on the symmetry axis $\mathcal{F}_1 = \mathcal{F}_2$.

\par The system of coupled PDEs resulting from the EsGB field equations is solved with the FIDISOL/CADSOL solver \cite{Solver1,Solver2,Solver3}, which implements a finite difference method together with the root finding Newton-Raphson method.  
We have also independently developed a new code to verify the  accuracy of solutions which utilizes pseudo-spectral methods (and which will be described in a forthcoming publication \cite{SpinningBHsCode}). The ADM mass M and angular momentum $J$ of the black hole solutions are read of the asymptotic decay of the metric functions
\begin{equation}
    g_{tt} = -1+\frac{2M}{r} + \mathcal{O}\left(r^{-2}\right), \qquad g_{t\varphi} = -\frac{2J}{r} \sin^2\theta + \mathcal{O}\left(r^{-2}\right).
\end{equation}
For future convenience we define the dimensionless spin of the solutions\footnote{For the case of a Kerr black hole, $0\leq \chi \leq 1$.}
\begin{equation}
    \chi \equiv J/M^2.
\end{equation}
Using the numerical method described above, in order to assess how the existence of a small mass limit changes with spin, we have explored the domain of existence of EsGB black holes for the linear, dilatonic, and quadratic exponential couplings.

\par
\begin{figure}[]
\centering
\includegraphics[width=0.5\textwidth]{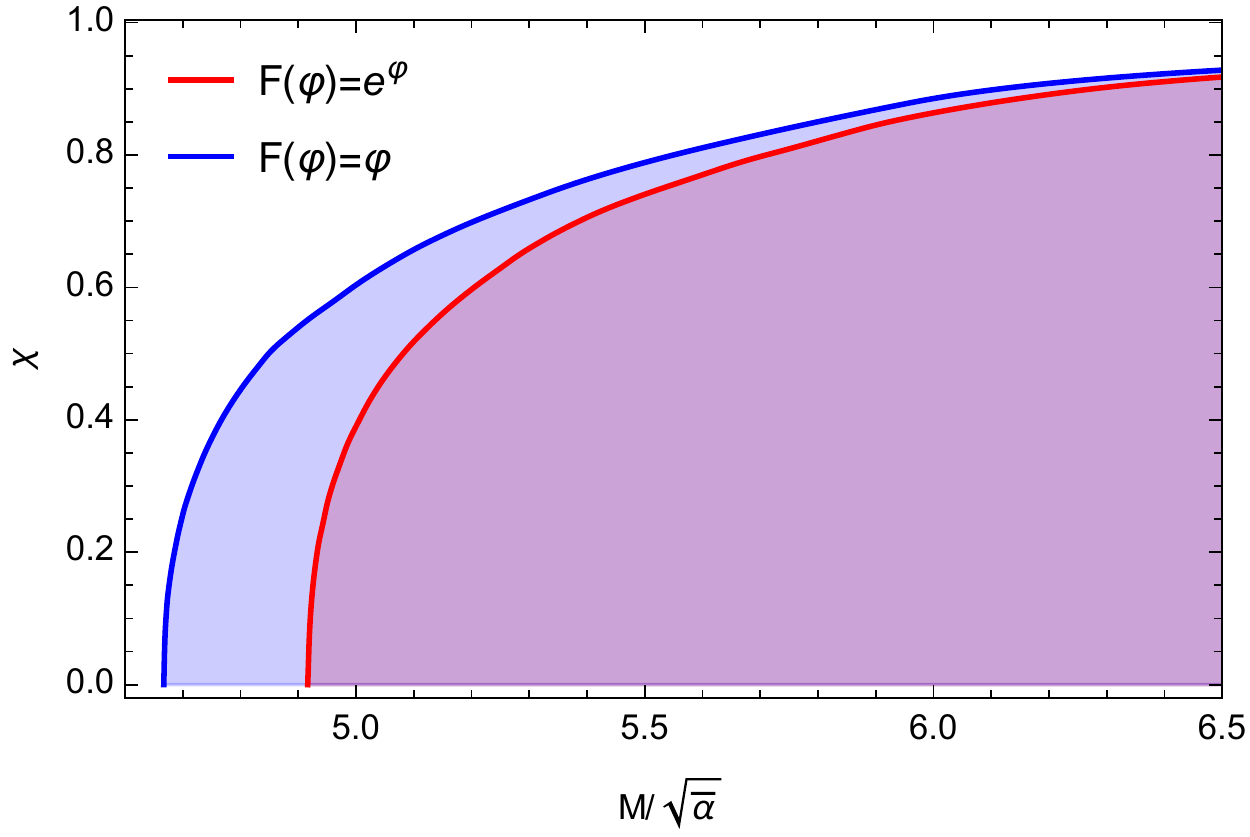}\hfill
\includegraphics[width=0.5\textwidth]{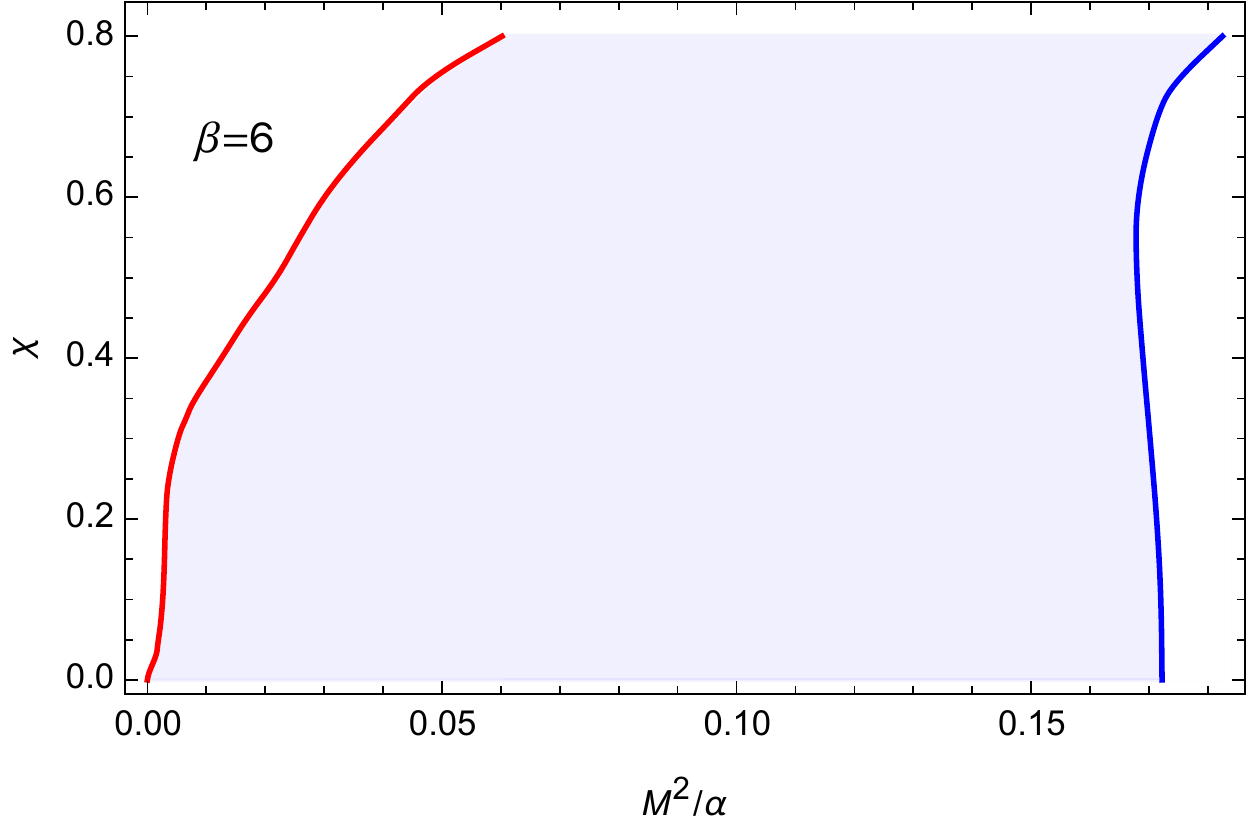}
\caption{(Left) Domain of existence of black hole solutions in the $\left(M/\sqrt{\overline{\alpha}},\chi\right)$ plane (shaded region), with dimensionless spins $\chi \lesssim 0.96$. The blue (red) line denotes the critical line for the linear (exponential) coupling. Note that here we used the convention of Eq. \eqref{eq:action_constraints}. (Right) Domain of existence of scalarized BH solutions in the $\left(M^2/\alpha,\chi\right)$ plane (shaded region), with dimensionless spins $\chi \lesssim 0.8$, for the coupling of Eq. \eqref{eq:coupling-quad} with $\beta=6$. The blue (red) line denotes the existence (critical) line. For small spins ($\chi \lesssim 0.1$), the absolute values that constitute the critical line should be taken with a pinch of salt because that region of the domain is particularly difficult to explore numerically as the mass of the solutions reach almost to zero.}
\label{fig:domain-linexp-spin}
\end{figure}

 For the dilatonic and linear couplings case ($F(\varphi) = e^\varphi$ and $F(\varphi) = \varphi$ in Eq. \eqref{eq:action_constraints}), the domain of existence of solutions with dimensionless spins $\chi \lesssim 0.96 $ is displayed in Fig. \ref{fig:domain-linexp-spin} (left). The domain of existence is bounded by a critical line. As we approached the critical line numerically, we observed a divergent behaviour on the Ricci and Gauss-Bonnet scalars e.g. on the equator on the horizon, and the code eventually crashed. We observe that higher spins result in higher values of the minimum allowed mass (for fixed coupling). This is not unexpected, as spin adds another repulsive effect to the system. In turn, this translates to tighter upper bounds on the allowed value of $\overline{\alpha}$, if spin is considered. Therefore, the constraints of Table \ref{tab:constraints} constitute legitimate upper bounds. From the point of view of cosmic censorship and Hawking evaporation, black holes in both the linear and dilatonic theories are expected to give rise to naked singularities as their endpoint, regardless of the initial spin of the solution, again raising questions about the consistency of these theories altogether.

We now consider the spontaneous scalarization coupling of Eq. \eqref{eq:coupling-quad} with the particular value $\beta=6$. In the static case we recall that no minimum mass was observed. We find, however, that when spin is brought into account, the picture changes and critical solutions do appear to exist, where curvature scalars diverge e.g. on the equator on the horizon, as observed in Fig. \ref{fig:domain-linexp-spin} (right). It is, however, unclear if the existence of rotating critical solutions changes the self consistency of the theory from the point of view of cosmic censorship and Hawking evaporation as it is not obvious that these solutions are ever reached. Indeed, a possibility is that during evaporation angular momentum is emitted at a (much) larger rate than mass, such that a rotating black hole spins down to a non-rotating state (which has no critical configuration) before most of its mass has been given up \cite{Page:1976ki}. The endpoint of evaporation for this coupling is therefore an open question and constitutes an avenue of further research.

Finally, we have obtained preliminary results for spin-induced scalarized black holes, exploring several values of $\beta$ (for the coupling of Eq. \eqref{eq:coupling-quad} with an overall reversed sign). In all cases, critical solutions were reached, in agreement with the results of Refs. \cite{Herdeiro:2020wei,Berti:2020kgk}.

\section{Discussion and conclusions}
\label{conclusions}
\par In this work we have explored the small mass limit of stationary (both static and spinning) black holes in theories containing Gauss-Bonnet terms in the action. Starting with an analytical example, we explored the small mass limit of black holes in the theory known as \textit{gravity with a generalized conformal scalar field} \cite{Fernandes:2021dsb}, which contains a Gauss-Bonnet term, and  where  static closed-form black hole solutions are known. These black holes do possess a minimum mass solution, where an inner singularity and the event horizon overlap. The inner singularity is intimately connected with the reality condition (that solutions must be real), because of the existence of terms containing square-roots on the solution (as is typical in Gauss-Bonnet theories). From a more physical point of view, the singularity is related to repulsive effects originating from the presence of the Gauss-Bonnet term in the theory.

\par Later, working with a more standard framework for EsGB theories, using numerical solutions of the field equations, a similar behavior was observed for the dilatonic \eqref{eq:coupling-exp} and linear couplings \eqref{eq:coupling-linear}. A curious case concerns the quadratic-exponential coupling \eqref{eq:coupling-quad}, where for sufficiently high values of the constant $\beta > \beta_{\mathrm{crit}}$ (defined in Eq. \eqref{eq:betacrit}), no static minimum mass solution was observed, thus showing that the existence of a critical singular black hole is not a generic prediction of theories containing Gauss-Bonnet terms. Then, from the point of view of cosmic censorship, this quadratic-exponential model might be viewed as more realistic option. The singularity structure for these models with $\beta > \beta_{\mathrm{crit}}$ is rather different from that of the dilatonic and linear and merits a deeper study. Once spin is considered, critical solutions do exist, but it is unclear if these are ever reached from Hawking evaporation. Also, for the coupling of Eq. \eqref{eq:coupling-quad}, scalarized black hole solutions exist only for curvatures above a certain threshold, rendering it particularly interesting.
\par Finally, we used the results concerning the minimum mass solutions into account to impose the tightest upper bounds to date on the coupling constant from observations, for both the dilatonic ($\sqrt{\overline{\alpha}} \lesssim \left(0.78 \pm 0.03\right)$ km) and linear ($\sqrt{\overline{\alpha}} \lesssim \left(0.82 \pm 0.03\right)$ km) theory, with the previous tightest upper bound being $\sqrt{\overline{\alpha}} \lesssim 1.18$ km \cite{Lyu:2022gdr}. Spin effects were found to only strengthen the previous upper limits.

\section*{Acknowledgements}

The authors thank Timothy Clifton for useful discussions and Nicola Franchini for comments on the manuscript. P. F. acknowledges support by the Royal Society grant RGF/EA/180022. D. J. M. is supported by a Royal Society University Research Fellowship. J. D. is supported by the Center for Research and Development in Mathematics and Applications (CIDMA) and Center for Astrophysics and Gravitation (CENTRA) through the Portuguese Foundation for Science and Technology (FCT - Fundação para a Ciência e a Tecnologia), references UIDB/04106/2020, UIDP/04106/2020 and UIDB/00099/2020. J. D. would like to also acknowledge the support from the projects PTDC/FIS-OUT/28407/2017, CERN/FISPAR/0027/2019, PTDC/FIS-AST/3041/2020 and from the European Union’s Horizon 2020 research and innovation (RISE) programme H2020-MSCA-RISE-2017 Grant No. FunFiCO-777740.

\appendix
\section{Onset of instability and spontaneous scalarization of a Schwarzschild black hole}
\label{app:onset}

Let us solve the perturbed scalar field equation \eqref{eq:perturbation} in a Schwarzschild spacetime background given by the line element of Eq. \eqref{eq:coupling_conditions} with
\begin{equation}
    f(r)=1-\frac{r_H}{r}, \qquad \delta(r)=0,
\end{equation}
where $r_H=2M$.
Taking into account that the background geometry is static and spherically symmetric, the scalar field perturbation can be separated in the following way
\begin{equation}
\delta \phi = \frac{u(r)}{r} e^{-i \omega t} Y_{\ell,m}(\theta, \varphi),
\end{equation}
where $Y_{\ell,m}(\theta, \varphi)$ are the spherical harmonics. The resulting equation for the radial part takes a Schrodinger-like form ($\ell=0$)
\begin{equation}
\frac{du}{dr^*} + \left(\omega^2 - V_{eff}\right) u = 0,
\end{equation}
where $dr^* = dr/f(r)$ and
\begin{equation}
V_{eff} = \left(1-\frac{r_H}{r}\right) \left(\frac{r_H}{r^3} - \frac{3 \alpha r_H^2}{2r^6} \right).
\end{equation}
A sufficient condition for the existence of an unstable mode (bound state as in quantum mechanics) is
\begin{equation}
\int_{-\infty}^{+\infty} V_{eff}(r^*)dr^* = \int_{r_H}^{+\infty} \frac{V_{eff}(r)}{1-\frac{r_H}{r}} dr < 0.
\end{equation}
The above condition gives $r_H/\sqrt{\alpha} < \sqrt{3/5} \approx 0.774597$ (or equivalently, $M/\sqrt{\alpha} \lesssim 0.387298$). Therefore Schwarzschild BHs with horizon radius obeying the previous condition, should be unstable in this framework. This, in turn, can be translated into a curvature condition: when the Gauss-Bonnet curvature at the horizon obeys
\begin{equation*}
\alpha^2 \mathcal{G}_{GR}|_{r_H} > \frac{100}{3},
\end{equation*}
the Schwarzschild BH should be unstable. This is only a sufficient condition for instability, but bifurcation of solutions actually occurs for slightly smaller curvature. To find the onset of instability we solve numerically the Schr\"odinger-like equation such that $\omega^2 = 0$ (when $\omega^2 < 0$ the tachyonic instability settles in). We find that the onset of instability occurs approximately at $r_H/\sqrt{\alpha} \approx 0.83$. This condition imposes a boundary on the domain of existence of (spontaneously) scalarized solutions. This result is valid for any coupling obeying the conditions of Eq. \eqref{eq:coupling_conditions}.

\section{EsGB Field Equations for a Static and Spherically Symmetric Background}
\label{app:feqs_static}
For a static and spherically symmetric background \eqref{eq:coupling_conditions}, the field equations take the form 
\begin{equation}
\mathcal{E}^{t}_{\phantom{t} t} = \frac{f \left(3 \alpha  f' \phi ' \dot \xi -\left(\phi '^2 \left(r^2-2 \alpha  (f-1) \ddot \xi\right)\right)+2 \alpha  (f-1) \phi '' \dot \xi -2\right)-f' \left(\alpha  \phi ' \dot \xi +2 r\right)+2}{2 r^2} = 0,
\end{equation}
\begin{equation}
\mathcal{E}^{r}_{\phantom{r} r} = \frac{\alpha  (3 f-1) \phi ' \left(f'-2 f \delta '\right) \dot \xi -2 r f'+r^2 f \phi '^2+f \left(4 r \delta '-2\right)+2}{2 r^2} = 0,
\end{equation}
\begin{equation}
\begin{aligned}
\mathcal{E}^{\theta}_{\phantom{\theta} \theta} = \mathcal{E}^{\varphi}_{\phantom{\varphi} \varphi} = \frac{1}{2r} \Bigg[& f \left(\alpha  f'' \phi ' \dot \xi -2 r \delta '^2+2 \delta '+2 r \delta ''-r \phi '^2\right)+f' \left(\delta ' \left(3 r-5 \alpha  f \phi ' \dot \xi \right)+\alpha  f \left(\phi '^2 \ddot \xi +\phi '' \dot \xi \right)-2\right) \\& + \alpha  f'^2 \phi ' \dot \xi -r f''+2 \alpha  f^2 \left(\phi ' \left(\left(\delta '^2-\delta ''\right) \dot \xi -\delta ' \phi ' \ddot \xi \right)-\delta ' \phi '' \dot \xi \right) \Bigg] = 0,
\end{aligned}
\end{equation}
and the scalar field equation is
\begin{equation}
\mathcal{E}_{\phi} = \frac{e^{\delta}}{r^2} \sbr{ \cbr{r^2 e^{-\delta} f \phi'}' + \frac{\alpha}{2}\dot \xi \cbr{\left(f-1\right)e^{\delta} \cbr{e^{-2\delta} f}'}' } = 0,
\end{equation}
where the primes denote a derivative with respect to $r$.

Once a closed-form expression for $\delta'$ is obtained by solving $\mathcal{E}^{r}_{\phantom{r} r}$, taking the $\mathcal{E}^{t}_{\phantom{t} t}$ and $\mathcal{E}_{\phi}$ equations, the above system can also be written in matrix form (Eq. \eqref{eq:system_mat}) with
\begin{equation}
    \begin{aligned}
        &\mathcal{M}_{11} = \frac{\alpha  (3 f-1) \phi' \dot \xi-2 r}{2 r^2}, \qquad \mathcal{M}_{12} = \frac{\alpha  (f-1) f \dot \xi}{r^2}\\&
        \mathcal{M}_{21} =  \frac{8 r^4 f \phi'+\alpha ^2 \phi' \left(f \left(r^2 (3 f (5 f-4)+1) \phi'^2+6 f (f+1)-14\right)+2\right) \dot \xi^2+2 \alpha  r \left(f \left(3 r^2 (1-3 f) \phi'^2-6 f+4\right)+2\right) \dot \xi}{4 r^2 f\left(\alpha  (1-3 f) \phi' \dot \xi+2 r\right)^2}\\&
        \mathcal{M}_{22} = \frac{f \left(\alpha  \dot \xi \left(4 r^3 (1-5 f) \phi'+\alpha  \left(r^2 (f (15 f-8)+1) \phi'^2-2 (f (3 f-7)+5)\right) \dot \xi\right)+8 r^4\right)+2 \alpha ^2 \dot \xi^2}{2 r^2 \left(\alpha  (1-3 f) \phi' \dot \xi+2 r\right)^2}\\&
        b_1 = \frac{f \left(\phi'^2 \left(r^2-2 \alpha  (f-1) \ddot \xi\right)+2\right)-2}{2 r^2}\\&
        b_2 = \frac{\alpha  \dot \xi \left(f \left(-2 \alpha  (f-1) (3 f-1) \phi'^2 \left(f \left(r^2 \phi'^2-2\right)+2\right) \ddot \xi+f \left(r^4 (5 f-1) \phi'^4+4 r^2 (21 f-5) \phi'^2-12 f+28\right)-20\right)+4\right)}{4 r^2 f \left(\alpha  (1-3 f) \phi' \dot \xi+2 r\right)^2}\\&
        + \frac{-4 r^3 f \phi' \left(f \left(r^2 \phi'^2+6\right)+2\right)-4 \alpha ^2 r f^2 (f (15 f-8)+1) \phi'^3 \dot \xi^2}{4 r^2 f \left(\alpha  (1-3 f) \phi' \dot \xi+2 r\right)^2}
    \end{aligned}
\end{equation}

The values appearing in Eq. \eqref{eq:det} are
\begin{equation}
    \begin{aligned}
    &\mathcal{A} = 3 \phi' \left(5 r^2 \phi'^2-4\right) \dot \xi^3,\qquad
    \mathcal{B} = -6 \dot \xi^2 \left(6 r^3 \phi'^2+\alpha  \phi' \left(r^2 \phi'^2-4\right) \dot \xi-2 r\right),\\&
    \mathcal{C} = \dot \xi \left(\alpha  \phi' \dot \xi+2 r\right) \left(14 r^3 \phi'-\alpha  \left(r^2 \phi'^2+12\right) \dot \xi\right),\qquad
    \mathcal{D} = -4 r \left(2 r^4+\alpha  \dot \xi \left(r^3 \phi'-3 \alpha  \dot \xi\right)\right).
    \end{aligned}
\end{equation}

\bibliography{biblio}

\end{document}